\documentclass[9pt,technote]{IEEEtran}

\usepackage{verbatim}

\usepackage{amssymb}
\usepackage{colortbl}
\usepackage[figuresright]{rotating}

\usepackage{placeins}
\usepackage{amsmath}
\usepackage
	{todonotes}
    
\usetikzlibrary{
calc,
backgrounds,
mindmap,
babel
}

\usepackage{
booktabs,
datetime,
graphicx,
multicol,
pgfplots,
ragged2e,
tabularx,
tikz,
wasysym
}

\usepackage{pgf}
\pgfplotsset{compat=1.8}

\usepackage{multirow}

\usepackage{listings}
\usepackage{xcolor}

\usetikzlibrary{arrows,shadows,petri}

\usepackage{hyperref}
\usepackage{xcolor}
\definecolor{darkblue}{rgb}{0, 0, 0.5}
\hypersetup{colorlinks=true,citecolor=darkblue, linkcolor=darkblue, urlcolor=darkblue}

\begin{document}

\title{Conversational Networks For Automatic Online Moderation}

\author{Etienne Papegnies\\ Vincent Labatut\\ Richard Dufour\\ Georges Linares%
\thanks{All authors are with \textit{Laboratoire Informatique d'Avignon} (LIA EA 4128), Avignon University, 84911 Avignon, France. E. Papegnies is additionally with \textit{Nectar de Code}, 13570 Barbentane, France. E-mail:  \{firstname.lastname\}@univ-avignon.fr}}



\maketitle

\begin{abstract}
Moderation of user-generated content in an online community is a challenge that has great socio-economical ramifications. However, the costs incurred by delegating this work to human agents are high. For this reason, an automatic system able to detect abuse in user-generated content is of great interest. There are a number of ways to tackle this problem, but the most commonly seen in practice are word filtering or regular expression matching. The main limitations are their vulnerability to intentional obfuscation on the part of the users, and their context-insensitive nature. Moreover, they are language-dependent and may require appropriate corpora for training.

In this paper, we propose a system for automatic abuse detection that completely disregards message content. We first extract a conversational network from raw chat logs and characterize it through topological measures. We then use these as features to train a classifier on our abuse detection task. We thoroughly assess our system on a dataset of user comments originating from a French Massively Multiplayer Online Game. We identify the most appropriate network extraction parameters and discuss the discriminative power of our features, relatively to their topological and temporal nature. Our method reaches an $F$-measure of $83.89$ when using the full feature set, improving on existing approaches. With a selection of the most discriminative features, we dramatically cut computing time while retaining most of the performance ($82.65$).
\end{abstract}


\vspace{0.2cm}
\textcolor{red}{\textbf{Cite as:} E. Papegnies, V. Labatut, R. Dufour \& G. linarès. ``\href{https://ieeexplore.ieee.org/document/8629298}{Conversational Networks For Automatic Online Moderation}'', IEEE Transactions on Computing Social Systems, in press, 2019. DOI: \href{https://doi.org/10.1109/TCSS.2018.2887240}{10.1109/TCSS.2018.2887240}}

\textcolor{red}{\textbf{Note:} the present preprint is a slightly longer version of this article.}

\section{Introduction}
\label{sec:Introduction}
Online communities have acquired an indisputable importance in today's society. From modest beginnings as places to trade ideas around specific topics, they have grown into important focuses of attention for companies to advertise products or governments interested in monitoring public discourse. They also have a strong social effect, by heavily impacting public and interpersonal communications.

However, the Internet grants a degree of anonymity, and because of that, online communities are often confronted with users exhibiting abusive behaviors. The notion of abuse varies depending on the community, but there is almost always a common core of rules stating that users should not personally attack others or discriminate them based on race, religion or sexual orientation. It can also include more community-specific aspects, e.g. not posting advertisement or external URLs. For community maintainers, it is often necessary to act on abusive behaviors: if they do not, abusive users can poison the community, make important community members leave, and, in some countries, trigger legal issues~\cite{FrenchLaw1, FrenchLaw2}. 
When users break the community rules, sanctions can then be applied. This process, called \textit{moderation}, is mainly done by humans. Since this manual work is expensive, companies have a vested interest in automating the process. 
%
In this work, we consider the classification problem consisting in automatically determining if a user message is abusive or not. 
This task is at the core of automated moderation, and it is difficult for several reasons.
First, the amount of noise in the content (typos, grammatical errors, uncommon abbreviations, out-of-vocabulary words...) of messages posted on the Internet is usually quite high.
Furthermore, as illustrated in Figure~\ref{fig:hard-noise}, while some of this noise is unwittingly produced by fast typing or poor language skills, a good part of it is voluntarily introduced as a means to defeat automated \textit{badword} checks.
\begin{figure}[htb]
	\centering
	\begin{lstlisting}
<User1> Pls d1e you f8 ck
	\end{lstlisting}
	\caption{Example of an abusive message with intentional noise.}
	\label{fig:hard-noise}
\end{figure}
Then, even with a noiseless message, it is sometimes necessary to perform advanced natural language analysis to detect abuse in a message. Figure~\ref{fig:hard-nl} gives a fictional example of a message containing no obvious indicators of abuse such as straight insults, while still being very abusive indeed.
\begin{figure}[htb]
	\centering 
	\begin{lstlisting}
<User2> Would you like to meet your maker? I
        can arrange that.
	\end{lstlisting}
	\caption{Example of a noiseless but abusive message.}
	\label{fig:hard-nl}
\end{figure}
Finally, even advanced natural language processing approaches may not be able to detect abuse from a message only without looking at its context. This context can take various forms. For instance, in the case of a ``Yo mama joke'', it is the continuation of the conversation. But it can also include external knowledge, which makes it harder to handle. For example, in Figure~\ref{fig:hard-context}, the message from \textit{User4} has no abuse markers at all until one considers both the messages that came before and historical knowledge.
\begin{figure}[htb]
\centering 
	\begin{lstlisting}
<User3> They've been discriminated against
        enough.
<User3> Six millions of them were killed 
        during the holocaust
<User4> @User3: that didn't actually happen.
	\end{lstlisting}
	\caption{Example of a message that is abusive only in a more general context.} 
	\label{fig:hard-context}
\end{figure}

To address these issues, we propose, as our main contribution in this paper, an approach that completely ignores the content of the messages and models conversations under the form of conversational graphs. By doing so, we aim to create a model that is not vulnerable to text-based obfuscation. We characterize these graphs through a number of topological measures which are then used as features, in order to train and test a classifier. Our second contribution is to apply our method to a corpus of chat logs originating from the community of the French massively multiplayer online game \textit{SpaceOrigin}\footnote{\url{https://play.spaceorigin.fr/}}. Our third contribution is to investigate the relative importance of the classification features, as well as the parameters of the graph extraction process, with regard to our classification task --the detection of abusive messages.

This paper is a significantly extended version of our preliminary work started in~\cite{Papegnies2017}. In comparison, we propose and experiment with several variations of our network extraction method, and vastly expand the array of features that we consider. We also adapt our approach to greatly increase the efficiency of our system with regard to necessary computational resources, and make it more versatile to possible use cases.

The rest of this paper is organized as follows. In Section~\ref{sec:Relwork}, we review related work on abuse detection and previous approaches dedicated to network extraction from various types of conversation logs. We describe the methods used throughout our pipeline in Section~\ref{sec:methods}, including the approach proposed to extract conversational networks, and the topological features that we compute to characterize them. In Section~\ref{sec:exp}, we present our dataset, as well as the overall experimental setup for the classification task. We then provide a discussion and a qualitative study of the performance of our approach, with a focus on the contributions of the considered features. Because some of them are computed from information that is not yet available at the instant some messages are posted, we also examine the performances of the system based only on information available at the time (\textit{i.e.} as a prediction task). Finally, we summarize our contributions in Section~\ref{sec:Conclusion} and present some perspectives for this work.

\section{Related Work}
\label{sec:Relwork}
In this section, we first review general approaches related to the problem of abuse detection (Section~\ref{sec:rel-abuse}), and then focus on techniques that have been previously used to extract graph-based representations of conversation logs (Section~\ref{sec:rel-network}).

\subsection{Abuse Detection}
\label{sec:rel-abuse}
One can distinguish two main categories of works related to abuse detection: those using the content of the targeted messages only, and those focusing on their context (user metadata, content of surrounding messages...). Some hybrid works also propose to combine both categories.

\paragraph{Content-Based Approaches}
The work initiated by Spertus in~\cite{Spertus1997} constitutes a first attempt to create a classifier for hostile messages. Abusive messages often contain hostility, so this task is related to ours. However, the notion of abuse is more general, as it can take a non-hostile form. Spertus uses static rules to extract linguistic markers for each message: Imperative Statement, Profanity, Condescension, Insult, Politeness and Praise. These are then used as features in a binary classifier. This approach obtains good results, except in specific cases like hostility through sarcasm. However, manually defining all the linguistic rules related to an abusive message is a severe limitation and appears impossible, in practice. Also, its application to another language would require to transpose it to other grammar rules and idioms.

Chen \textit{et al}.~\cite{Chen2012q} seek to detect offensive language in social media so that it can be filtered out to protect adolescents. Like before, this task is more specific than ours, as using offensive language is just one type of abuse. Chen \textit{et al}. developed a system that uses lexical and syntactical features as well as user modeling, to predict the offensiveness value of a comment. They note that the presence of a word tagged as \textit{offensive} in a message is not a definite indication that the message itself is offensive. For instance, while ``You are stupid'' is clearly offensive, ``This is stupid xD'' is not. They further show that lack of context can be somewhat mitigated by looking at word $n$-grams instead of unigrams (\textit{i.e.} single words). The method relies on manually constituted language-dependent resources though, such as a lexicon of offensive terms, which also makes it difficult to transpose to another language.

Dinakar \textit{et al}.~\cite{Dinakar2011} use $tf$--$idf$ features, a static list of badwords, and of widely used sentences containing verbal abuse, to detect cyberbullying in Youtube comments. Bullying is mainly characterized by its persistent and repetitive nature, and it can therefore be considered as a very specific type of abuse. Like before, the proposed model shows good results except when sarcasm is used. It is worth noting that sarcasm can be considered as a form of natural language obfuscation that is especially hard to detect in written communications, because of the lack of inflexion clues. 

In \cite{Chavan2015}, Chavan \& Shylaja review machine learning (ML) approaches to detect cyberbullying in online social networks. They show that pronoun occurrences, usually neglected in text classification, are very important to detect online bullying. They use skip-gram features to mitigate the sentence-level context issues by taking into account distant words. These new features allow them to boost the accuracy of a classifier detecting bullying by $4$ percent points. The approach is however still vulnerable to involuntary misspellings and word-level obfuscation. It uses a language-dependent list of badwords during preprocessing.

In their recent article, Mubarak \textit{et al}.~\cite{Mubarak2017} work on the detection of offensive language in Arabic media, by introducing the interesting possibility of dynamically generating and expanding a list of bad words. They extract a corpus of tweets that is divided into two classes (\textit{obscene} / \textit{not obscene}) based on static rules. They then perform a log odds ratio analysis to detect the words favoring documents from the obscene class. Such an approach could be very useful in an online classification setting, but inherently requires a dataset where the number of samples in the obscene class is large. Still, they show that a list of words dynamically generated using that method contains $60\%$ of new obscene words, and the process can be iterated over. Relatively to our problem of interest, the main limitation of this article is its focus on obscene words, which are just one specific type of abuse.

In \cite{Razavi2010}, Razavi \textit{et al}. focus on a wider spectrum of types of abuse than the previously cited works, which they call \textit{inflammatory comments}. It ranges from impoliteness to insult, and includes rants and taunts. To detect them, they stack three levels of Naive-Bayes classifier variants, fed with features related to the presence, frequency, and strength of offensive expressions. These are computed based on a manually constituted lexicon of offensive expressions and insults, which makes the method relatively corpus-specific. The resulting system shows high precision and has the useful characteristics of being updatable online. It is however vulnerable to the text-based obfuscation techniques we have previously mentioned.

With recent developments in GPU architecture and hardware availability, more computationally expansive techniques have been used. In~\cite{Djuric2015}, Djuric \textit{et al}. detect hate speech in Twitter data. They adopt a two-step approach consisting in first learning a low-dimensional representation of the tweets, and then applying a classifier to discriminate them based on this representation. They note that jointly using message- and word-embeddings instead of simple bag-of-words boosts the performance. Park \& Fung~\cite{Park2017} also work on tweets using neural networks, but they focus only on sexism- and racism-related cases. They propose a two-step framework consisting in first training a Convolutional Neural Network (CNN) to identify the absence/presence of abuse, and then performing a simple logistic regression to further discriminate between sexism and racism. Both of these approaches are inherently portable, however they require a lot of data. 

In \cite{Pavlopoulos2017}, Pavlopoulos \textit{et al}. develop an automatic moderation system for comments posted by users on Websites. It is based on a Recurrent Neural Network (RNN) operating on word-embeddings, with an attention mechanism. They apply it to two large corpus extracted from a Greek sports website and the English Wikipedia. The proposed system outperforms CNN and other more mainstream classifiers. However, it is worth noticing that these tasks are slightly different, as the the Greek corpus is annotated for general moderation, whereas the English one focuses on personal attacks.

It is worth noting that all these ML-based approaches perform better when a large dataset is available for training. However text-based approaches are usually language-dependent, which means that models have to be trained on a dataset of the specific language. This is usually not an issue when classifying English messages because of the wealth of publicly-available data, but is problematic in our case, since our messages come from low resource language communities.

Content-based text classification usually makes for a good baseline. However, such methods have severe limitations. First, abuse can be spread over a succession of messages. Some messages can even reference a shared history between two users. Second, it is very common for users to voluntarily obfuscate message content to work around badwords detection. Indeed, abusers can bypass automatic systems by making the abusive content difficult to detect: for instance, they can intentionally modify the spelling of a forbidden word.

Hosseini \textit{et al}. \cite{hosseini2017} demonstrate such an attack against the Google Perspective API\footnote{\url{https://www.perspectiveapi.com}}. Adversarial attacks based on word-level obfuscation are nothing new, and approaches exist to counter them. For instance, Lee \textit{et al}.~\cite{Lee2005} experiment with spam de-obfuscation using a Hidden Markov Model that incorporates lexical information. Such an approach yields good results for de-obfuscation, but it is computationally expensive and requires a dataset of obfuscated words for training. More recently, Rojas \textit{et al}.~\cite{Rojas2017} describe a more compact approach based on a dynamic programming sequence alignment algorithm. It has a different set of limitations, the main one being that it does not allow for one character to be used as an obfuscated version of several distinct original characters (it uses a one-to-one character mapping).

\paragraph{Context-Based Approaches}
Because the reactions of other users to an abuse case are completely beyond the control of the abuser, some works consider the content of messages \textit{around} the targeted message, instead of the content of the targeted message only. 

For instance, Yin \textit{et al}. \cite{Yin2009} use features derived from the sentences neighboring a given message to detect harassment on the Web. Harassment implies repetition, and can be considered as a specific type of abuse. Their goal is to spot conversations going off-topic, and use that as an indicator. Their combined content/context approach shows good results when used against multi-participant chat logs. They also note that sentiment features seem to constitute mostly noise due to the high misspelling rate. This lack of discriminative power from sentiment features is something we have also noticed while experimenting with content-based techniques on our data in~\cite{Papegnies2017a}.

In \cite{Cheng2015a}, Cheng \textit{et al}. do not try to perform automatic moderation. Instead, they conduct a comprehensive study of antisocial behavior in online discussion communities, and use its results to build user behavior models. We include this work in our review, because it provides some insight into the devolution of abusive users over time in a community, regarding both the quality of their contributions and their reactions towards other members of the community. A critical result of this analysis is that instances of antisocial messages usually generate a bigger response from the community, compared to normal messages. In our own work, we build upon this observation and compare classification performances obtained when considering or ignoring messages published right after the classified message.

Balci \& Salah in~\cite{Balci2015} take advantage of user features to detect abuse in the community of an online game. These features include information such as gender, number of friends, financial investment, avatars, and general rankings. The goal is to help human moderators dealing with abuse reports, and the approach yields sufficiently good results to achieve this. One important difference with our work is that in our case, the user data necessary to replicate this approach are not available. As a practical consideration the availability of that data will always depend on the type of the community.

In our own previous work~\cite{Papegnies2017a}, we tackle the same problem as in this article, \textit{i.e.} detect abuse in chat messages in the context of an online game. However, unlike the method proposed presently, we use a wide array of language features (bag-of-words, $tf$-$idf$ scores, sentiment scores...) as well as context features derived from the language models of other users. We also experiment with several advanced preprocessing approaches. This method allows us to reach a performance of $72.1\%$ in terms of $F$-measure on our abusive message detection task.

Of all the approaches of the literature described in this section, only our previous work~\cite{Papegnies2017a} as well as Balci \& Salah's~\cite{Balci2015} aim at solving the same problem as us. The others focus on tasks which are related to abuse detection, but still different, and generally more specific, e.g. insult or cyberbullying detection. The work of Balci \& Salah differs from ours in the way they solve the problem, as they focus on the users' profiles and behaviors: these data are not available in our case, so we only use the published messages. Our previous work~\cite{Papegnies2017a} is completely based on the textual content of the messages, whereas the one presented here ignores it, and relies only on a graph-based modeling of the conversation, which is completely new in this context. Another important methodological difference with the literature is that almost all content-based methods rely on manually constituted linguistic resources, which makes them difficult to transpose to another context (different language or online community). By comparison, our present approach is completely language independent, as it does not use the textual content (apart from user names). The third difference is that almost all methods from the literature consider messages independently, when we use sequences of messages forming conversations. Finally, we use a classic classifier to determine if a message is abusive, which means that our approach requires much less training data than the deep learning methods that we mentioned earlier.


\subsection{Network Extraction from Conversation Logs}
\label{sec:rel-network}
Although a major part of the methods proposed to address the abuse detection problem focus on the content of the exchanged messages, it appears that a user with previous exposure to automatic moderation techniques can easily circumvent them~\cite{hosseini2017}. To avoid this issue, a solution would be not to focus on the textual content, but rather on the interactions between the users through these messages. For instance, the number of respondents to a given message appears frequently as a classification feature in the literature, \textit{e.g.} as in~\cite{Cheng2015a}. But graphs constitute a more natural paradigm to model such relational information, under the form of so-called \textit{conversational networks}, which represent the flow of the conversation between users. Such networks have the advantage of including the mentioned feature (number of respondents), but also much more information regarding the way users interact. We adopted this approach in our previous work~\cite{Papegnies2017}, which is the first attempt at using such graph-based conversation models to solve a general abuse detection problem. Our present work is an extension of this method, essentially on two aspects: we experiment with several variations of our graph-extraction process, and we consider much more graph-based features.

This section reviews methods proposed in the literature for the extraction of conversational networks. We do not narrow it to the abuse detection context, as our previous work \cite{Papegnies2017} would be the only one concerned. Even so, there are not many works dealing with the extraction of conversational networks. This may be due to the fact that the task can be far from trivial, depending on the nature of the available raw data: it is much harder for chat logs than for structured messages board or Web forums, for instance. In a multi-participant chat log it is frequent to see multiple disjointed conversations overlapping. There is no fixed topic although some chatrooms have a general purpose. There is also no built-in mechanism to specify the message someone is responding to. Finally, in most IRC (Internet Relay Chat) chat logs, there is no enforcement mechanism to ensure that users have only one nickname.

In~\cite{Mutton2004}, Mutton proposes a strategy to extract such a network from IRC chat logs. The goal is to build a tool to visualize user interactions in an IRC chat room over time. The author uses a simple set of rules based on \textit{direct referencing} (\textit{i.e.} when a user addresses another one by using his nickname), as well as temporal proximity and temporal density of the messages. In our own work, we adapt and expend on some of these rules, whereas certain cannot be applied. Specifically, while in a regular IRC channel timestamps are indeed useful to determine intended recipients of a message, in our case they are basically irrelevant.

Osesina \textit{et al}. in \cite{Osesina2012} build on the work of Mutton using Response Time Analysis, which assumes that both temporal proximity and the cyclical nature of conversations can be used to perform edge prediction. The authors also use the content of the communications to build a word network, and then assign edges between users based on the keywords they use and the presence of these keywords in word clusters. Finally, by combining these two approaches with direct addressing, they achieve impressive performance in edge prediction with regard to a manually extracted network, both in terms of edge existence and edge strength. It is worth noticing the significant computational requirement for large chat logs. Besides the targeted task itself, the main difference with our approach is that this one is strongly content-based.

In~\cite{Gruzd2008}, Gruzd \textit{et al}. push the usage of direct referencing further by developing methods of name discovery. The data they work on come from a bulletin board which shares some similarities with regular chat: linear stream of messages with possibly intertwined discussion threads. By comparing a network extracted through their name discovery method, to a chain network based on temporal proximity, they show that their approach is better suited to detect social network links. Useful takeaways of their method are: the use of neighboring words (for instance, \textit{Dr.}, \textit{Pr.}, \textit{Jr}, are often seen in proximity of person names, whereas \textit{Street}, \textit{Ave} are often near location names), capitalization, and the position of words within the document (\textit{e.g.} their sample of posts often end with a user's signature because a bulletin board does not have the ephemeral nature of chatrooms). However, these difference between the two media also makes this method unsuitable to our data.

Camtepe \textit{et al.} in \cite{Camtepe2004} experiment on the detection of groups of users in chat logs, collected from three different chatrooms in the Undernet IRC network. They first build a matrix containing the numbers of messages posted by each user at each considered time step. It can be considered as a low-resolution view of the logs --it retains information about temporal proximity but looses sequential information. They then perform Singular-Value-Decomposition (SVD) on this matrix, in order to ease the identification of clusters of interacting users (\textit{i.e.} conversations). They extract an approximation of the conversational network from this partition, by representing each cluster by a clique. They validate their approach by manually extracting the actual conversational network directly from the logs, and comparing their structures. The main difference with our situation is that the conversational graph is only seen as a way to validate the user group detection method: we want to use it as a model of the interactions.

Forestier \textit{et al}. in \cite{Forestier2011} tackle the extraction of networks from online forums. While the structure of conversations is explicitly represented on certain platforms, this it not the case there: a thread is represented as a flat sequence of messages. This makes it challenging to determine the intended recipient of a message. The authors show that, by using a combination of grammatical analysis and Levenshtein distance computation for substrings, they can often ascertain who talks to whom. The resulting network can then be used to analyze the role of users in the community. The main difference with our method is that we ignore the content of messages.

Travassoli \textit{et al}. in \cite{Tavassoli2014} explore different methods to extract representative networks from group psychotherapy chat logs. One of them includes fuzzy referencing to mitigate effects of misspelled nicknames, and rules for representing one-to-all messages. The bulk of the methods uses static patterns of exchanges to predict a receiver. Their system shows a good agreement score with a human annotator. It is worth noting, though, that these logs are substantially different from ours: the psychotherapy sessions have well defined boundaries and a limited number of participants. This prevent the transposition of this approach to our problem.

Sinha \& Rajasingh in \cite{Sinha2014a} use only direct referencing, but with the same fuzzy matching strategy as in~\cite{Tavassoli2014}, in order to extract a network representing the activity in the \texttt{\#ubuntu} IRC support channel. This method manages to expose high level components of the Ubuntu social network, which in turn allows for the qualification of user behaviors into specific classes such as \textit{beginner} or \textit{expert}. This method of building user models can be very interesting when the data describing the users are scarce, as is the case on IRC where everyone can join and there is no requirement to register. While it does not allow for the direct classification of individual messages, the behavior information can be useful as a supporting feature in a text classification task.

Anwar \& Abulaish in \cite{Anwar2014} build a framework allowing to query user groups and communities of interest, based on the data extracted from the computers of suspects during a criminal investigation. They use a social graph extraction method that relies both on the presence of communication between users and the overlap between the content of the messages they exchange, in order to assign weights to the edges of the network. They then experiment with various forms of community detection (\textit{i.e.} graph partitioning) to identify groups of users in this network. However, this method assumes that the corpus contains a variety of topics allowing to discriminate the groups, which is not necessarily the case for us, since our logs are thematically dominated by the video game hosting the chatroom.

An interesting task where conversational networks can be used is the detection of controversial discussions. In \cite{Garimella2015}, Garimella \textit{et al}. show that the predefined types of interactions allowed by Twitter can be used to build networks that highlight the presence of polarized groups of users. They extract all tweets matching a given hashtag around the time a specific event happens, then detect an endorsement link thanks to Twitter's retweet feature. The resulting graph is then partitioned and analyzed using a controversy measure. In our context it is difficult to adopt this approach, as endorsement information is not immediately available and would have to be inferred from message content.

The methods proposed in the literature mainly rely on the content of the exchanged messages. By comparison, our method only focuses on the presence/absence of communication between the users, \textit{i.e.} on the dynamics of the conversation and its structure. Some methods also rely on specific functionalities of the studied platforms (\textit{e.g.} answers explicitly addressed to a user), which are absent from our own data.

\section{Methods}
\label{sec:methods}
In this section, we describe the methods that we propose to compute the features later used in the classification task to separate abusive and non-abusive messages. We first present how we extract conversational networks from series of raw chat messages (Section~\ref{sec:methods-netext}), before describing the topological measures that we use to characterize these networks (Section~\ref{sec:methods-features}). 

\subsection{Network Extraction}
\label{sec:methods-netext}
We extract networks representing conversations between users through a textual discussion channel. They take the form of weighted graphs, in which the vertices and edges represent the users and the communication between them, respectively. An edge weight corresponds to a score estimating the intensity of the communication between both connected users. We propose two variants of our method, allowing to extract \textit{undirected} vs. \textit{directed} networks. In the latter case, the edge direction represents the information flow between the considered users. Note that each network is defined relatively to a \textit{targeted message}, since the goal of this operation is to provide features used to classify the said message.

The method that we use to extract the networks representing the conversations in which each message occurs has three steps, that we describe in detail in this section. First, we identify the subset of messages that we will use to extract the network (Section~\ref{sec:net-context}). Second, we select as nodes a subset of users which are likely receivers of each individual message (Section~\ref{sec:net-slide}). Third, we add edges and revise their weights depending on the potential receivers (Section~\ref{sec:net-distro}). We describe and discuss the resulting conversational graphs in Section~\ref{sec:net-example}

\subsubsection{Context period}
\label{sec:net-context}
\begin{figure}[!ht]
    \center
    \resizebox{\linewidth}{!}{
	\begin{tikzpicture}[scale=0.45]
	\definecolor{lightblue}{RGB}{220, 220, 255}
	\definecolor{lightgreen}{RGB}{220, 255, 220}
	
	\foreach \x in {1,1.5,...,21}{\draw[] (\x,0) rectangle (\x+0.5,2);}
	\foreach \x in {1.5,2,...,20.5}{\draw[fill=lightgray] (\x,0) rectangle (\x+0.5,2);}
	
	\draw[fill=red] (11,0) rectangle (11.5,2);
	\node[] at (11.25,5.75) {Targeted message};
	\draw (11.25,5.25) -- (11.25,2);
	
	\draw[decorate,decoration={brace,mirror,amplitude=10pt},yshift=-30pt] (1.5,-0.5) -- (21,-0.5) node [midway,yshift=-0.6cm]{Context period};
	\draw[decorate,decoration={brace,amplitude=10pt},yshift=-30pt] (1.5,4.5) -- (11,4.5) node [midway,yshift=0.6cm]{Past messages};
	\draw[decorate,decoration={brace,amplitude=10pt},yshift=-30pt] (11.5,4.5) -- (21,4.5) node [midway,yshift=0.6cm]{Future messages};
	
	\foreach \x in {1.5,21}{\draw[dotted] (\x,-1.65) -- (\x,3.60);}
	\foreach \x in {11,11.5}{\draw[dotted] (\x,2.00) -- (\x,3.60);}
	\draw[ultra thick,loosely dotted] (-0.5,1) -- (0.8,1);
	\draw[ultra thick,loosely dotted,->] (21.7,1) -- (23.3,1);
	
	\draw[fill=blue, thick] (7.5,0) rectangle (8,2);
	\node[] at (7.75,-1.10) {Current message};
	\draw (7.75,0) -- (7.75,-0.75);
	
	\node[] at (5.25,2.75) {Sliding window};	
	\draw[ultra thick] (2.5,-0.25) rectangle (8,2.25);
 \end{tikzpicture}
 
	}
    \caption{Sequence of messages (represented by vertical rectangles) illustrating the various concepts used in our conversational network extraction process. Figure available at \href{https://doi.org/10.6084/m9.figshare.7442273.v3}{10.6084/m9.figshare.7442273} under CC-BY license.}
	\label{fig:cs_vs_ws}
\end{figure}
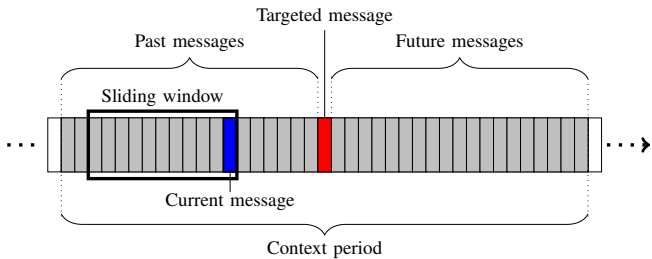

Our first step is to determine which messages to use in order to extract the network. For this purpose, we define the \textit{context period}, as a sequence of messages. Figure~\ref{fig:cs_vs_ws} shows an example of context period, representing each message as a vertical rectangle. Note that time flows from left to right in the figure. This sequence is centered on the \textit{targeted message} (in red), and spans symmetrically \textit{before} (left side) and \textit{after} (right side) its occurrence. Put differently: we consider the same number of past and future messages. The networks extracted from the context period contain only the vertices representing the users which posted at least once on this channel, during this period.

Besides the network extracted over the whole context period (before and after the targeted message), which we call the \textit{Full} network, we also consider two additional networks. We split the period in the middle, right on the targeted message, and extract one network over the messages published in the first half (\textit{Past messages}), called \textit{Before} network, and one over the other half (\textit{Future messages}), called \textit{After} network. Both of those smaller networks also contain the targeted message. For a prediction task, \textit{i.e.} when using only past information to classify the targeted message, one would only be able to use the \textit{Before} network. However, in a more general setting, all three networks (\textit{Before}, \textit{After} and \textit{Full}) can be used.

\subsubsection{Sliding Window}
\label{sec:net-slide}
In order to extract a network, we apply an iterative process, consisting in sliding a window over the whole context period, one message at a time, and updating the edges and their weights according to the process described next. The size of this \textit{sliding window} is expressed in terms of number of messages, and it is fixed. It can be viewed as a focus on a part of the conversation taking place at some given time. It is shown as a thick black frame in Figure~\ref{fig:cs_vs_ws}. We call \textit{current message} the \textit{last} message of the window taken at a given time (represented in blue), and \textit{current author} the author of the current message.

The use of such a fixed-length sliding window is a methodological choice justified by four properties of the user interface of the considered discussion channel: 1) At any given time, the user can see only up to $10$ preceding messages without scrolling; 2) when a user joins a channel, the server sends him only the last $20$ messages posted on the channel; 3) it is impossible for a user to scroll back the history further than $20$ lines; and 4) the user interface masks \textit{join} and \textit{part} events by default, whereas in typical chat clients the arrival and departure of users are shown by default. Thus, at some given time, a user only has access to a limited knowledge regarding who is participating in the conversation. As explained later, we use this value of $20$ messages as an upper bound, and experiment with different sliding window sizes.

\subsubsection{Weight Assignment}
\label{sec:net-distro}
Our assumption is that the current message is destined to the authors of the other messages present in the considered sliding window. Based on this hypothesis, we update the edges and weights in the following way. We start by listing the authors of the messages currently present in the sliding window, and ordering them by their last posted message. Only the edges towards the users in that list will receive weight. This choice is also due to the user interface constraints: \textit{a priori}, a user cannot reliably know which users will receive a given message. Furthermore, the data we have do not allow us to directly determine channel occupancy at the time a message is posted. 

\begin{figure}[!ht]
    \centering
    \resizebox{\linewidth}{!}{
	\begin{tikzpicture}[scale=0.45]
			\foreach \x in {3.5,4,...,9.5}{\draw[fill=lightgray] (\x,0) rectangle (\x+0.5,2);}
			\foreach \x in {4,5.5,8,9}{\draw[fill=blue] (\x,0) rectangle (\x+0.5,2);}
			\foreach \x in {6,7,8.5}{\draw[fill=green] (\x,0) rectangle (\x+0.5,2);}
			\foreach \x in {5,7.5}{\draw[fill=orange] (\x,0) rectangle (\x+0.5,2);}
			\foreach \x in {4.5,6.5}{\draw[fill=purple] (\x,0) rectangle (\x+0.5,2);}
			\draw[fill=blue, very thick] (9,0) rectangle (9.5,2);
			
			\node[] at (6.75,2.75) {\scriptsize Sliding window};	
			\draw[ultra thick] (4,-0.25) rectangle (9.5,2.25);
			
            \draw[thick] (10.4,-0.3) -- (10.4,3.25);
            
			\node[above left] at (12.00,2.5) {\scriptsize Rank};	
			\node[] at (11.5,2.25) {1.};	
			\node[] at (11.5,1.50) {2.};	
			\node[] at (11.5,0.75) {3.};	
			\node[] at (11.5,0.00) {4.};	
            
			\draw[fill=purple, draw=lightgray] (9.10,1.60) rectangle (9.40,1.90);
			\draw[fill=cyan, draw=lightgray] (9.10,1.20) rectangle (9.40,1.50);
			
			\node[anchor=south] at (12.75,2.50) {$a$};
			\draw[fill=blue]   (12.5, 2.00) rectangle (13.0,2.50);
			\draw[fill=green]  (12.5, 1.25) rectangle (13.0,1.75);
			\draw[fill=orange] (12.5, 0.50) rectangle (13.0,1.00);
			\draw[fill=purple] (12.5,-0.25) rectangle (13.0,0.25);
            
			\node[anchor=south] at (14.00,2.50) {$b$};
			\draw[fill=green]  (13.75,2.00) rectangle (14.25,2.50);
			\draw[fill=orange] (13.75,1.25) rectangle (14.25,1.75);
			\draw[fill=purple] (13.75,0.50) rectangle (14.25,1.00);			
            
			\node[anchor=south] at (15.625,2.50) {$c$};
			\draw[fill=purple] (15.00,2.00) rectangle (15.50,2.50);
			\draw[fill=cyan]   (15.75,2.00) rectangle (16.25,2.50);
			\draw[fill=green]  (15.375,1.25) rectangle (15.875,1.75);
			\draw[fill=orange] (15.375,0.50) rectangle (15.875,1.00);			

			\node[above right] at (16.50,2.5) {\scriptsize Weight};	
			\node[anchor=west] at (16.75,2.25) {+++};	
			\node[anchor=west] at (16.75,1.50) {++};	
			\node[anchor=west] at (16.75,0.75) {+};	
\end{tikzpicture}
    }
    \caption{Example of sliding window (left) and computation of the corresponding receivers' scores (right). Each color represents a specific user. On the left, each message in the window is filled with the color of its author, whereas the small squares represent direct references to users. On the right, the $a$, $b$, $c$ columns represent the different steps of the computation (see text). Figure available at \href{https://doi.org/10.6084/m9.figshare.7442273.v3}{10.6084/m9.figshare.7442273} under CC-BY license.}
	\label{fig:weight}
\end{figure}
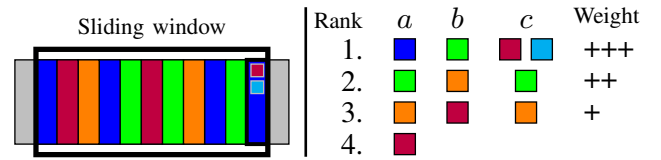

The left part of Figure~\ref{fig:weight} displays an example of sliding window, in which the colors of the messages (vertical rectangles) represent their authors. So, in this specific case, $4$ different users participate in the conversation. Ordered from latest to earliest, these are: blue (author of the current message, \textit{i.e.} the rightmost in the window), green, orange and red. This list of users is noted $a$ in the right part of the figure. Obviously, a user is not writing to himself, so we remove the current author from the list, resulting in list $b$. The use of such an ordered list is justified by the assumption of \textit{Temporal Proximity}, which appears commonly in the literature concerned with the extraction of conversational networks (\cite{Mutton2004, Osesina2012, Camtepe2004}). It states that the most recent a message, the most likely its author to be the recipient of the current message. 

The user interface allows us to \textit{explicitly} mention users in a message by their name, and moreover the game prevents the users from changing their name: we need to take these properties into account. It is also a common assumption that the presence of direct referencing increases the likelihood that the referred person is the intended recipient of the message. To reflect this in our process, we move the users directly referenced in the \textit{current message} at the top of the list. If some users are directly referenced although they have not posted any message in the considered window, they are simply \textit{inserted} at the top of the list. In Figure~\ref{fig:weight}, direct references are represented as small colored squares located in the current message. There are two of them in our example, referring to the purple and cyan users. The former has one post in the window, so he is moved from the third to the first rank in the list. The latter did not post anything in the window, so he is inserted at the first position. This results in what we call the \textit{list of receivers}, which appears as list $c$ in the figure.

We now want to connect the current author to the receivers constituting our ordered list. Our choice to create or update edges towards all users in the window even in case of direct referencing is based on several considerations. First, directly referencing a user does not imply that he is part of the conversation or that the message is directed towards him: for instance, his name could just be mentioned as an object of the sentence. Second, there can be multiple direct references in a single message (as in our example). Third, in online public discourse, directly addressing someone does not mean he is the sole intended recipient of the message. For instance when discussing politics, a question directed towards someone can have as a secondary objective to have the target expose his stance on an issue to the other participants.

We also want to adjust the strength of each of these connections depending on the rank of the concerned receivers: the higher the rank, the stronger the interaction. For this purpose, each receiver is assigned a score, which is a decreasing function of both his rank $i$ in the list and of the length $N$ of this list (as reflected by the number of $+$ signs in Figure~\ref{fig:weight}). We propose three different scoring functions, defined so that the assigned weights sum to unity:
\begin{itemize}
	\item \textbf{Uniform:} Each receiver gets the same weight, defined as:
	\begin{equation}
		f_U(i) = \frac{1}{N}
	\end{equation}
	\item \textbf{Linear:} The score decreases as a linear function of the rank:
	\begin{equation}
		f_L(i) = \frac{N-i}{\sum_{j=1}^{N}j}
	\end{equation}
	\item \textbf{Recursive:} The first receiver gets $60\%$ of the total weight, and the rest of them share the remaining $40\%$ using the same recursive $60$--$40\%$ split scheme:
	\begin{equation}
       f_R(i) = \begin{cases} 
       			0.6 \times 0.4^{i-1}, & \text{if } 1 \leq i < N \\
                0.4^{i-1}, & \text{if } i = N
              \end{cases}
	\end{equation}
\end{itemize}

\begin{figure}[!ht]
    \centering
	    \includegraphics[width=0.99\linewidth]{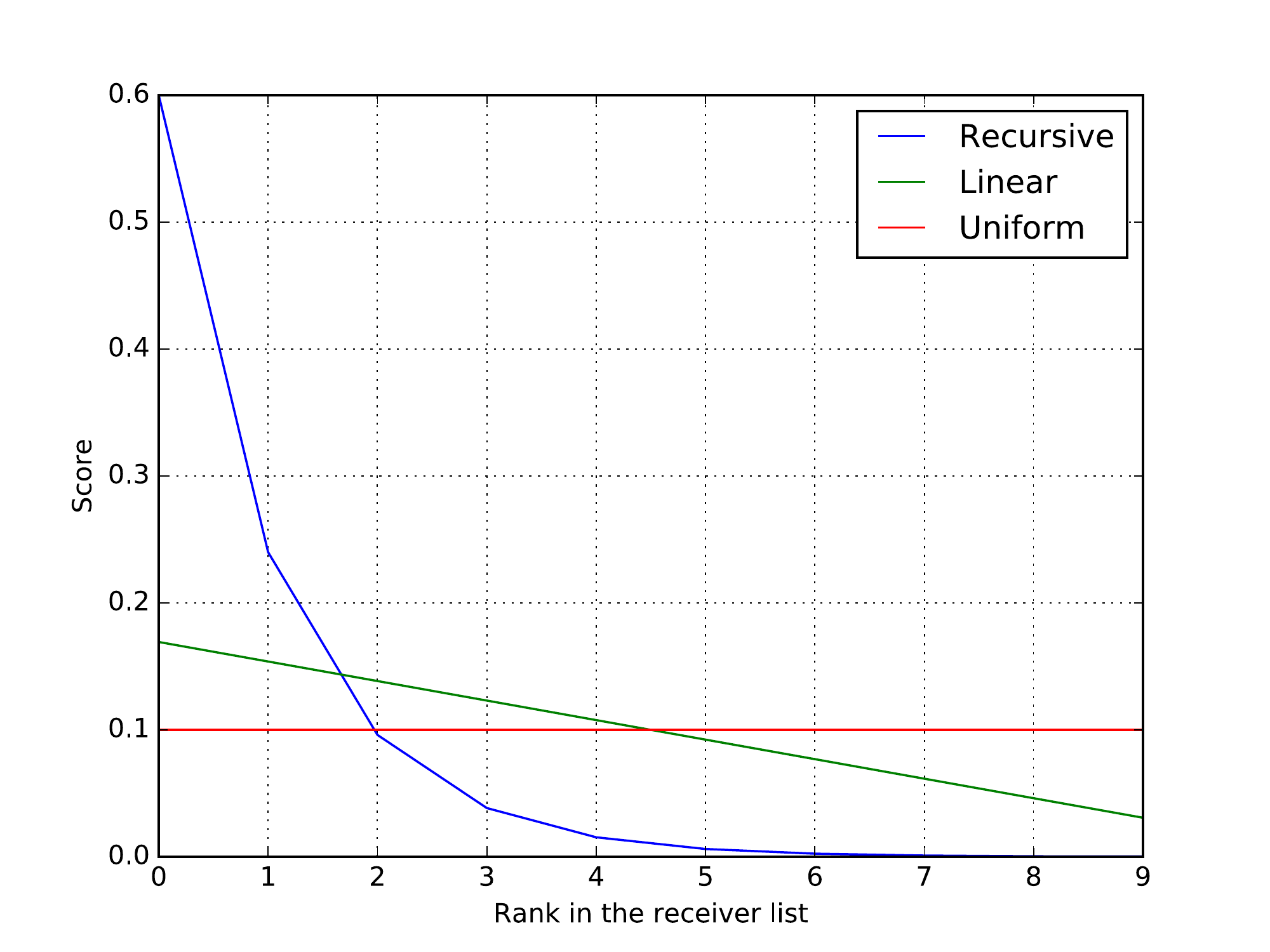}
    
    \caption{Scores assigned by our $3$ scoring functions $f_U$, $f_L$ and $f_R$ for a receiver list containing $10$ users.}
	\label{fig:distribs}
\end{figure}

As an illustration, Figure~\ref{fig:distribs} displays the scores assigned by these three strategies for $N = 10$, as functions of the receiver's rank. The \textit{Uniform} strategy $f_U$ (in red) assumes that the content of the communication is not really important, and that the goal of the current author is just to have the message seen by as much people as possible. It therefore places very little importance on temporal proximity or direct referencing. The \textit{Recursive} approach $F_R$ (in blue) gives the most importance to direct referencing and temporal proximity, with scores dropping fast when the receiver is not directly referenced or the author of the immediately preceding message. Finally, the \textit{Linear} approach $f_L$ (in green) also places the most importance on temporal proximity and direct referencing, but in a less contrasted way, since it assigns higher scores (compared to $f_R$) to receivers located at the bottom of the list. We later compare these $3$ strategies during our experiments, in order to determine whether it is worth exploring more advanced scoring functions, or if the difference in performance is not significant enough to justify this.

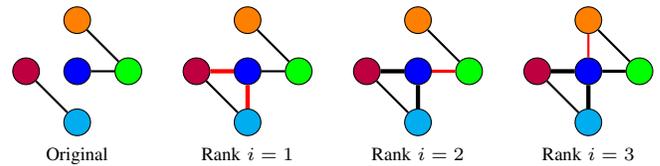
\begin{figure}[!ht]
    \center
    \resizebox{\linewidth}{!}{
	\begin{tikzpicture}[scale=0.45]
			\node[circle, draw=black, fill=green, minimum size=10pt] (G1) at (6.0,-4) {};
			\node[circle, draw=black, fill=purple, minimum size=10pt] (P1) at (3.0,-4) {};
			\node[circle, draw=black, fill=cyan, minimum size=10pt] (C1) at (4.5,-5.5) {};
			\node[circle, draw=black, fill=orange, minimum size=10pt] (O1) at (4.5,-2.5) {};
			\node[circle, draw=black, fill=blue, minimum size=10pt] (B1) at (4.5,-4) {};
			\draw[thick] (B1) -- (G1);
			\draw[thick] (G1) -- (O1);
			\draw[thick] (C1) -- (P1);
           	\node[below] at (4.50,-6.00) {\scriptsize Original};	
            
			\node[circle, draw=black, fill=green, minimum size=10pt] (G2) at (11.0,-4) {};
			\node[circle, draw=black, fill=purple, minimum size=10pt] (P2) at (8.0,-4) {};
			\node[circle, draw=black, fill=cyan, minimum size=10pt] (C2) at (9.5,-5.5) {};
			\node[circle, draw=black, fill=orange, minimum size=10pt] (O2) at (9.5,-2.5) {};
			\node[circle, draw=black, fill=blue, minimum size=10pt] (B2) at (9.5,-4) {};
			\draw[thick] (B2) -- (G2);
			\draw[thick] (G2) -- (O2);
			\draw[thick] (C2) -- (P2);
			\draw[ultra thick, draw=red] (B2) -- (P2);
			\draw[ultra thick, draw=red] (B2) -- (C2);
			\node[below] at (9.50,-6.00) {\scriptsize Rank $i=1$};	
           
			\node[circle, draw=black, fill=green, minimum size=10pt] (G3) at (16.0,-4) {};
			\node[circle, draw=black, fill=purple, minimum size=10pt] (P3) at (13.0,-4) {};
			\node[circle, draw=black, fill=cyan, minimum size=10pt] (C3) at (14.5,-5.5) {};
			\node[circle, draw=black, fill=orange, minimum size=10pt] (O3) at (14.5,-2.5) {};
			\node[circle, draw=black, fill=blue, minimum size=10pt] (B3) at (14.5,-4) {};
			\draw[thick] (G3) -- (O3);
			\draw[thick] (C3) -- (P3);
			\draw[ultra thick] (B3) -- (P3);
			\draw[ultra thick] (B3) -- (C3);
			\draw[very thick, draw=red] (B3) -- (G3);
			\node[below] at (14.50,-6.00) {\scriptsize Rank $i=2$};	
            
			\node[circle, draw=black, fill=green, minimum size=10pt] (G4) at (21.0,-4) {};
			\node[circle, draw=black, fill=purple, minimum size=10pt] (P4) at (18.0,-4) {};
			\node[circle, draw=black, fill=cyan, minimum size=10pt] (C4) at (19.5,-5.5) {};
			\node[circle, draw=black, fill=orange, minimum size=10pt] (O4) at (19.5,-2.5) {};
			\node[circle, draw=black, fill=blue, minimum size=10pt] (B4) at (19.5,-4) {};
			\draw[thick] (G4) -- (O4);
			\draw[thick] (C4) -- (P4);
			\draw[ultra thick] (B4) -- (P4);
			\draw[ultra thick] (B4) -- (C4);
			\draw[very thick] (B4) -- (G4);
			\draw[thick, draw=red] (B4) -- (O4);
			\node[below] at (19.50,-6.00) {\scriptsize Rank $i=3$};	
			
\end{tikzpicture}
    }
    \caption{Update of the edges and weights of the conversational graph corresponding to our ongoing example. The first graph displays the state before the update, and each remaining one corresponds to one rank in the receiver list. Figure available at \href{https://doi.org/10.6084/m9.figshare.7442273.v3}{10.6084/m9.figshare.7442273} under CC-BY license.}
	\label{fig:build}
\end{figure}

We can then update the graph by creating an edge between the current author and each user in the receiver list. We consider two possible approaches, leading to an \textit{undirected} vs. a \textit{directed} network. In the latter case, the edge is directed from the current author towards the receiver, in order to model the communication flow. Each newly created edge is assigned a weight corresponding to the receiver's score. If this edge already exists, we increase its current weight by the said score. Figure~\ref{fig:build} shows the result of this update based on our previous example from Figure~\ref{fig:weight}, for the extraction of an undirected network. The first graph represents the network before the update. It already contains some edges though, resulting from some previous processing. The remaining graphs of the figure represent the changes corresponding to the ranks appearing in the receiver list: first position (purple and cyan users), second position (green) and third position (orange). Red edges represent the edges being modified or created. If we were extracting a directed graph, then the new edges would be directed outward from the central blue vertex.

Once the iterative process has been applied for the whole context period, we get what we call the \textit{Full} network. As mentioned before, for testing matters we also process $2$ lesser networks based on the same context: the \textit{Before} and \textit{After} networks are extracted using only the messages preceding and following the \textit{targeted message}, respectively, as well as the \textit{targeted message} itself.

\subsubsection{Extracted Networks}
\label{sec:net-example}
\begin{figure*}[!ht]
	\center
	\resizebox{\linewidth}{!}{
		\includegraphics[width=0.333\textwidth]{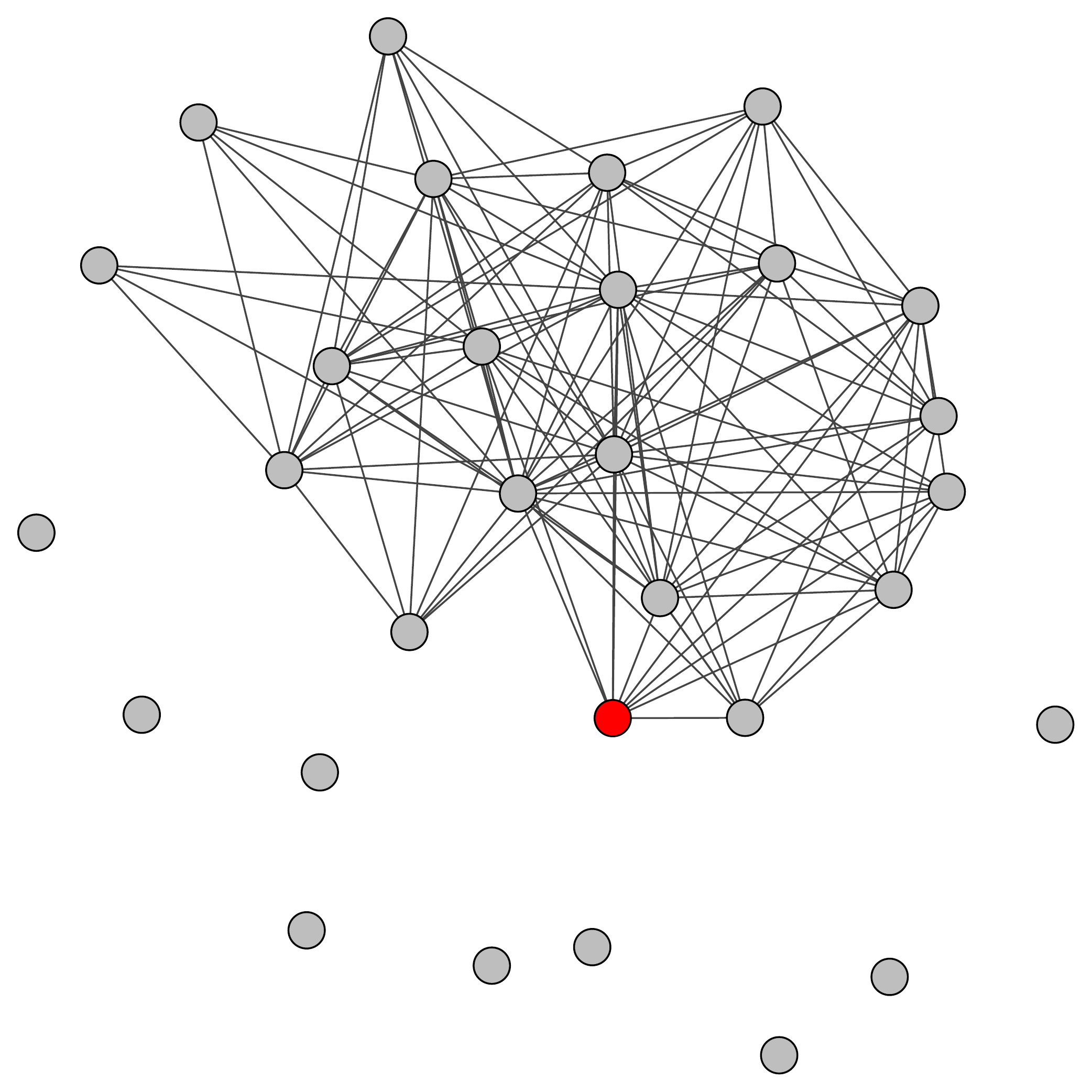}\hfill
		\includegraphics[width=0.333\textwidth]{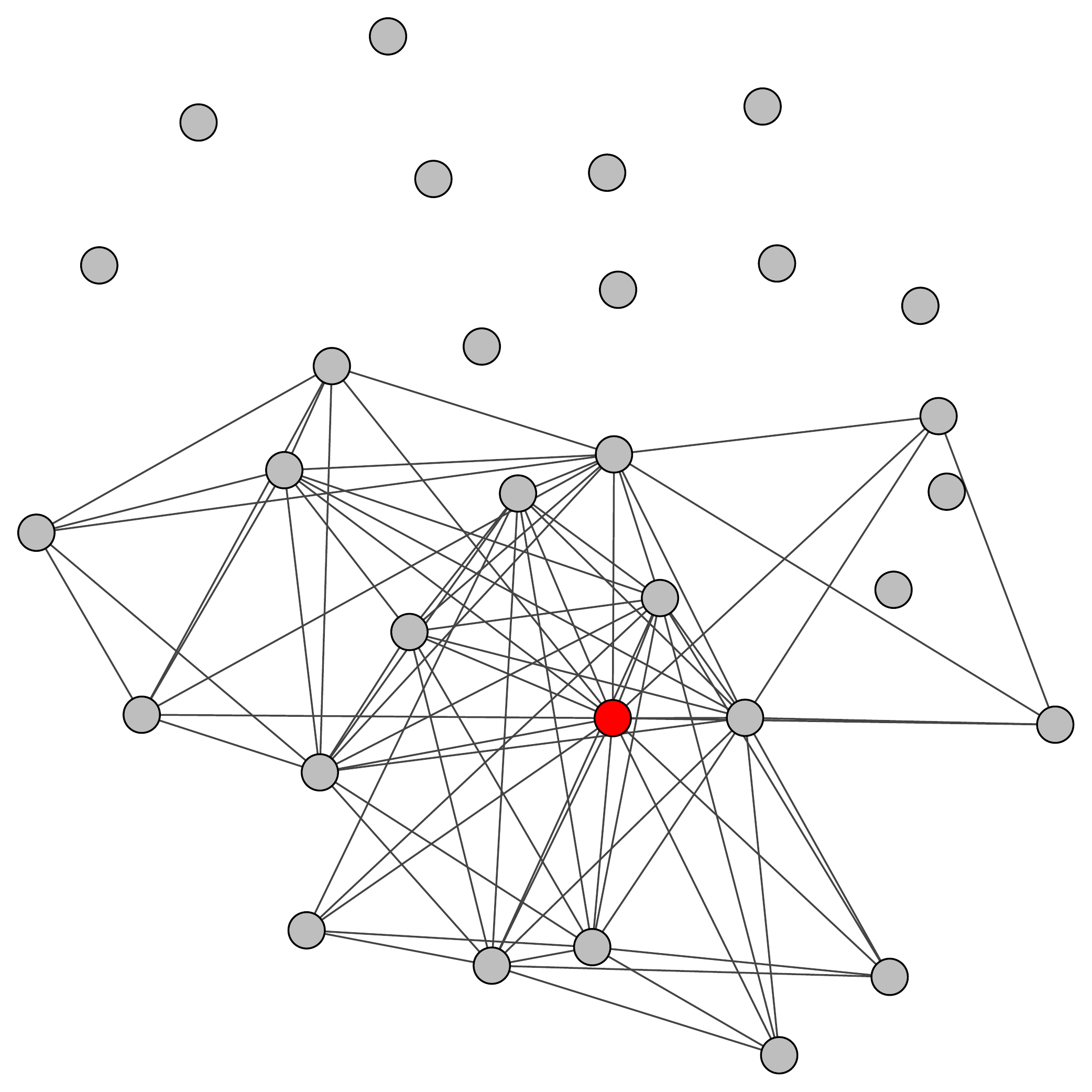}\hfill
		\includegraphics[width=0.333\textwidth]{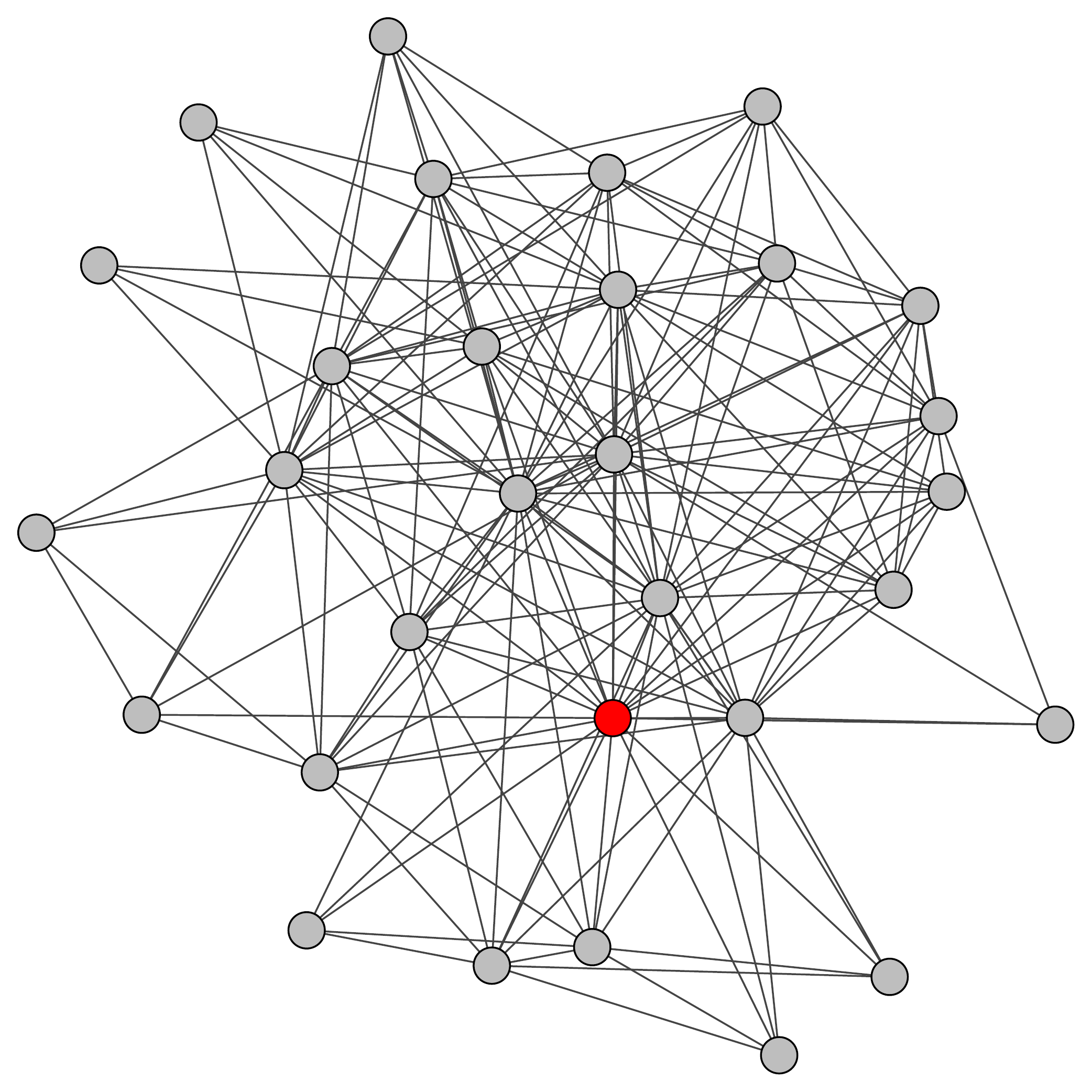}
	}
	\caption{Example of the $3$ types of conversational networks extracted for a given context period: \textit{Before} (left), \textit{After} (center), and \textit{Full} (right). The author of the targeted message is represented in red. For readability reasons, weights and directions have been omitted. Figure available at \href{https://doi.org/10.6084/m9.figshare.7442273.v3}{10.6084/m9.figshare.7442273} under CC-BY license.}
	\label{fig:nets}
\end{figure*}

Figure~\ref{fig:nets} shows a real-world example of the three conversational networks obtained by applying our extraction method to an abusive comment belonging to our dataset. They are obtained based on a context period of $200$ messages, a sliding window of $10$ messages, and are undirected. The isolates (disconnected vertices) present in the \textit{Before} and \textit{After} networks correspond to users present in the context period, but active only after or before the targeted message, respectively. The red vertex corresponds to the author of the targeted message, which we call the \textit{targeted user}. One can see that the users involved in the conversation, as well as the location of the targeted user in this conversation, undergo some dramatic changes after the abuse.

Generally speaking, two vertices are connected in our networks if they are supposed to have a direct interaction. Thus, if only one conversation occurs during the considered context period, we expect the network to be rather cliquish. It seems possible to have several communities, \textit{i.e.} several loosely connected dense subgraphs, if certain users completely ignore some other ones, for some reason. However, the smoothing induced by our use of a sliding window is likely to hide this type of behavior, especially if the window is large. The presence of a community structure could also occur if several distinct conversations take place during the considered context period. However, this can happen only if the number of common users between the conversations is small compared to the network size (otherwise, the communities will be indistinguishable). Due to the relatively dense nature of the networks (when ignoring isolates), we think weights are likely to be an important information, allowing to separate accidental edges from relevant ones. The edge direction allows distinguishing unilateral and bilateral interactions, so it could help identify certain types of conversations with atypical structure (\textit{e.g.} one-way communication).

\subsection{Features}
\label{sec:methods-features}
The classification features that we consider in this work are all based on topological measures allowing to characterize graphs in various ways. We process all the features for each of the $3$ types of networks (\textit{Before}, \textit{After}, and \textit{Full}) described in Section~\ref{sec:methods-netext}.

\begin{table}
	\centering
    \caption{List of the features used to characterize the conversational networks. The letters in the \textit{Wght.} (Weights) and \textit{Dir.} (Directions) columns stand for: \textit{Unweighted} or \textit{Undirected} (U), \textit{Weighted} (W), \textit{Directed} (D), \textit{Incoming} (I) and \textit{Outgoing} (O). In addition, each vertex-centered feature is averaged over the vertex set $V$ to get a corresponding graph-centered feature.}
    \begin{tabular}{|l|l|l|l|l|l|}
        \hline
    	Scale & Scope & Name & Ref. & Wght. & Dir. \\
        
        \hline
        \multirow{15}{*}{Graph} & \multirow{7}{*}{Macro} & Weak Components & -- & -- & U \\
        & & Strong Components & -- & -- & D \\
        & & Adhesion/Cohesion & \cite{White2001} & -- & D \\
        & & Articulation Points & -- & -- & U \\
        & & Diameter & -- & U/W & U/D \\
        & & Radius & -- & U & U/I/O \\
        & & Average Distance & -- & U & U/D \\
        \cline{2-6}
        & \multirow{3}{*}{Meso} & Clique Count & -- & -- & -- \\
        & & Communities & -- & U & D \\
        & & Modularity & \cite{Newman2004e} & U/W & U \\
        \cline{2-6}
        & \multirow{5}{*}{Micro} & Edges/Vertices & -- & -- & -- \\
        & & Density & -- & -- & -- \\
        & & Global Transitivity & \cite{Luce1949} & U & U \\
        & & Reciprocity & \cite{Wasserman1994} & -- & D \\
		& & Degree Assortativity & \cite{Newman2002} & -- & U/D \\
		
		\hline
        \multirow{20}{*}{Vertex} & \multirow{10}{*}{Macro} & Eigenvector Centrality & \cite{Bonacich1972} & U/W & U/D \\
        & & Hub/Authority Scores & \cite{Kleinberg1999} & U/W & D \\
        & & Alpha Centrality & \cite{Katz1953} & U/W & D \\
        & & Power Centrality & \cite{Bonacich1987} & U & D \\
        & & PageRank Centrality & \cite{Brin1998} & U/W & U/D \\
        & & Subgraph Centrality & \cite{Estrada2005} & U & U \\
        & & Betweenness Centrality & \cite{Freeman1977} & U/W & U/D \\
        & & Closeness Centrality & \cite{Bavelas1950} & U/W & U/I/O \\
        & & Eccentricity & \cite{Harary1969} & U & U/I/O \\
        & & Articulation Point & \cite{Harary1969} & -- & U \\
        \cline{2-6}
        & \multirow{6}{*}{Meso} & Coreness Score & \cite{Seidman1983} & -- & U/I/O \\
        & & Participation Coefficient & \cite{Guimera2005a} & U & U/I/O \\
        & & Internal Intensity & \cite{Guimera2005a} & U & U/I/O \\
        & & External Intensity & \cite{Dugue2014a} & U & U/I/O \\
        & & Diversity & \cite{Dugue2014a} & U & U/I/O \\
        & & Heterogeneity & \cite{Dugue2014a} & U & U/I/O \\
        \cline{2-6}
        & \multirow{4}{*}{Micro} & Degree Centrality & \cite{Shaw1954} & U & U/I/O \\
        & & Strength Centrality & \cite{Barrat2004} & W & U/I/O \\
        & & Local Transitivity & \cite{Watts1998} & U/W & U \\
        & & Burt's Constraint & \cite{Burt2004} & U/W & -- \\
        \hline
	\end{tabular}
    \label{tab:FeatureList}
\end{table}

We adopt an exploratory approach and consider a large range of topological measures, focusing on the most widespread in the literature. Our selection is listed in Table~\ref{tab:FeatureList}. Some of these measures can optionally handle edge directions or edge weights: we consider all practically available variants, in order to assess how informative these aspects of the graph are relatively to our classification problem. 

One can distinguish topological measures in terms of scale and scope. The \textit{scale} depends on the nature of the characterized entity: vertex, subgraph or graph. In our case, we focus only on \textit{vertex-} and \textit{graph-focused} measures: the former allows focusing on the author of the targeted message, whereas the latter describes the whole conversation, but we do not have any subgraph to characterize. The \textit{scope} corresponds to the nature of the information used to characterize the entity: \textit{microscopic} (interconnection between a vertex and its direct neighborhood), \textit{mesoscopic} (structure of a subgraph and its direct neighborhood), and \textit{macroscopic} (structure of the whole graph).

In the rest of this section, we describe these measures briefly: first the vertex-focused ones (Section~\ref{sec:NodeMeasures}), then the graph-focused ones (Section~\ref{sec:GraphMeasures}). For each measure, we give a generic, graph-theoretical definition, before explaining how it can be interpreted in the context of our conversational networks.

\subsubsection{Vertex-Focused Topological Measures}
\label{sec:NodeMeasures}
These measures allows characterizing only a single vertex. We compute them all for the vertex corresponding to the author of the targeted message (represented in red in Figure~\ref{fig:nets}).

\paragraph{Microscopic Measures} 
We start with the measures which describe a vertex depending on its direct neighborhood. In our context, this amounts to characterizing the position of some user depending on its direct interlocutors. In the case of a conversation involving a very small number of persons, it is likely all of them interact directly, and so these measures can also help describing the conversation itself.

The \textit{Degree Centrality} is a normalized version of the standard degree~\cite{Shaw1954, Freeman1978}, which corresponds itself to the number of direct neighbors of the considered vertex. In a directed graph, one can distinguish an incoming and an outgoing Degree Centrality, focusing only on the incoming and outgoing edges of the vertex, respectively. In our case, it can be interpreted as the number of users that have exchanged (undirected version), received (outgoing) or sent (incoming) messages to the author, respectively. We use both undirected and directed variants of the Degree Centrality.

The generalization of the degree to weighted networks is called the strength~\cite{Barrat2004}. The \textit{Strength Centrality} is based on the sum of the weights of the edges attached to the considered vertex. Like the Degree, it is possible to use incoming and outgoing versions if the network is directed. In our conversational graph, compared to the degree, the strength takes into account the frequency of the interactions. This allows accounting for certain situations ignored by the Degree Centrality. For instance, a user can have a few interlocutors, but still be central if he exchanges a lot with them. We use both undirected and directed variants of the \textit{Strength Centrality}.

The \textit{Local Transitivity} (or \textit{Clustering Coefficient})~\cite{Watts1998} corresponds to the proportion of edges between the considered vertex's neighbors, relatively to what this number could be if all of them were interconnected. It ranges from $0$ (no inter-neighbor edge at all) to $1$ (the vertex and its neighborhood form a clique). In our context, a high transitivity indicates that the user belongs to a single conversation, in which most protagonists exchange messages. On the contrary, a low transitivity denotes some form of segmentation: either the user participates in several distinct conversations, or some of his interlocutors ignore each other. We use the unweighted original version and the weighted variant presented in~\cite{Barrat2004}. 

\textit{Burt's Constraint} \cite{Burt2004} measures how redundant the neighbors of the vertex of interest are. 
It is based on the idea that a vertex located at the interface between several independent groups holds a position of power. Burt's Constraints measures this level of independence through a non-linear combination of the number of connections between the neighbors. A high value indicates how embedded the vertex is in its neighborhood. In our case, this can help distinguishing users depending on the number of conversations they are involved in, if we suppose a conversation corresponds to a clique-like structure. We use both unweighted and weighted variants of Burt's Constraint.

\paragraph{Macroscopic Measures}
The measures harnessing the entirety of the graph structure form the largest group. In our context, they allow characterizing the position of a vertex relatively to the whole context period (\textit{Full}) or to one of its halves (\textit{Before} and \textit{After}).

So-called \textit{spectral measures} are based on the spectrum of the graph adjacency matrix, or of a related matrix. The \textit{Eigenvector Centrality} \cite{Bonacich1972} can be considered as a generalization of the degree, in which instead of just counting the neighbors, one also takes into account their own centrality: a central neighbor increases the centrality of the vertex of interest more than a peripheral one. Central vertices tend to be embedded in dense subgraphs. We use the (un)weighted and (un)directed variants of the measure (so: $4$ variants in total). 

One limitation of the Eigenvector Centrality is that if the graph is directed and not strongly connected, certain vertices systematically get a zero centrality, whatever their position. Several modifications have been proposed to handle this situation. The \textit{Hub} and \textit{Authority Scores} \cite{Kleinberg1999} are two complementary measures processed through the HITS algorithm (Hyperlink-Induced Topic Search). They solve the issue by splitting the centrality value into two parts: one for the incoming influence (Authority), and the other for the outgoing one (Hub). We use the (un)weighted directed variants of both Hub and Authority scores. 

The \textit{Alpha Centrality} (or \textit{Katz Centrality}) \cite{Katz1953, Bonacich2001} solves the same problem by assigning a minimal positive centrality value to all vertices. Additionally, it allows attenuating the influence of distant vertices during the computation. We use the (un)weighted directed variants of this measure.
The \textit{Power Centrality} \cite{Bonacich1987} generalizes both the Eigenvector and Alpha Centralities. In particular, it allows a negative attenuation. The implementation we use only works for unweighted directed graphs.

The \textit{PageRank Centrality} \cite{Brin1998} can be seen as a variant of the Katz Centrality. One limitation of the later is that when a central vertex has many outgoing edges, all of them receive all its influence, as if they were its only recipient. The PageRank Centrality includes a normalization allowing to model the dilution of this influence. We use the (un)weighted and (un)directed variants of this measure.

Compared to the other spectral measures, the \textit{SubGraph Centrality} \cite{Estrada2005} defines the notion of reachability based on closed walks rather than simple walks. Put differently, the other spectral measures consider that the vertex of interest influences (resp. is influenced by) some other vertex if a walk exists to go to (resp. come from) this vertex. The Subgraph Centrality requires both, and it uses an attenuation coefficient to give less importance to longer walks. The implementation we use only deals with undirected unweighted graphs.

In our conversational graph, we expect that a user participating a lot in the conversation will be central, and even more so if there are several conversations and he is participating in the main one. It is difficult to predict which ones of these slightly different spectral measures will be the most appropriate to our case, which is why we included all of the available ones.

\vspace{0.3cm}
Another group of macroscopic measures is based on the notions of shortest path or geodesic distance (\textit{i.e.} the length of the shortest path). 

The \textit{Betweenness Centrality} \cite{Freeman1977} is related to the number of shortest paths going through the considered vertex. In communication networks such as ours, it can be interpreted as the level of control that the user of interest has over information transmission. We use the (un)weighted and/or (un)directed variants of this measure.

The \textit{Closeness Centrality} \cite{Bavelas1950} is related to the reciprocal of the total geodesic distance between the vertex of interest and the other vertices. It is generally considered that it measures the efficiency of the vertex to spread a message over the graph, and its independence from the other vertices in terms of communication. The \textit{Eccentricity} \cite{Harary1969} is related to the Closeness Centrality, but it is not a centrality measure. On the contrary, it quantifies how \textit{peripheral} the vertex of interest is, by considering the distance to its farthest vertex. By comparison to the Closeness Centrality, there is no reciprocal involved, and it uses the maximum operator instead of the sum. In our case, both measures indicate how involved the considered user is in the conversation(s), as they directly depend on how directly connected he is to the other users. In particular, we expect important changes in the \textit{Before} and \textit{After} graphs to reflect a significant modification of the user's role in the conversation. For the Closeness Centrality, we use the (un)weighted and (un)directed variants, but for the Eccentricity, we only have access to the unweighted (un)directed variants.

\vspace{0.3cm}
The last group of macroscopic measures is based on the notion of connectivity, \textit{i.e.} whether or not a path exists between certain parts of the graph. 

An \textit{Articulation Point} (or \textit{Cut Vertex}) is a vertex whose removal makes the graph disconnected, \textit{i.e.} split it into several separate components~\cite{Harary1969}. We define a binary nodal feature indicating if the vertex of interest is an articulation point ($1$) or not ($0$). It could help describing whether the targeted user is bridging two separate groups of users in the conversation, possibly indicating that he caused a topic shift or that some of the users have left the conversation.

\paragraph{Mesoscopic Measures} 
Mesoscopic measures rely on an intermediate structure to characterize a vertex. In our case, such a subgraph corresponds to a tightly knit group of users, and is likely to represent a conversation. So, this type of measure would allow characterizing the position of a vertex relatively to the various conversations taking place in the considered context period (provided there are several of them).

The \textit{Coreness Score}~\cite{Seidman1983} is based on the notion of $k$-core, which is a maximal induced subgraph whose all vertices have a degree of at least $k$. The Coreness Score of a vertex is the $k$ value of the $k$-core of maximal degree to which it belongs. In our context, the Coreness Score is related to the number of participants of the largest conversation involving the user of interest. We use an undirected version of the Coreness Score, as well as two variants focusing on incoming and outgoing edges in directed networks.

We also take advantage of the \textit{Within-Module Degree} and \textit{Participation Coefficient}, a pair of complementary measures defined relatively to the community structure of the graph~\cite{Guimera2005a}. We detect the community structure through the InfoMap method~\cite{Rosvall2008}. These measures aim at characterizing the position of a vertex at this intermediate level. The \textit{Within-Module Degree} (or \textit{Internal Intensity}) assesses the internal connectivity. It evaluates how the degree of a vertex within his community relates to those of the other vertices from the same community. For us, it is an indicator of how involved the user is in his current conversation. The \textit{Participation Coefficient} is concerned with the external connectivity: it is based on the number and quality of the connections that the vertex has outside of his own community. In our case, a high value could indicate either someone holding a mediation position, in the case of a single conversation involving several groups of users, or someone participating in several conversations. We use the original undirected variants of these measures, as well as the directed variants proposed in~\cite{Dugue2015a} to focus on incoming and outgoing edges.

One limitation of the Participation Coefficient is that it mixes several aspects of the external connectivity: the number of external connections, the number of concerned external communities, and the distribution of these connections over these communities. To solve this issue, three measures were proposed in \cite{Dugue2014a} to separately assess these three properties. They are respectively called \textit{External Intensity}, \textit{Diversity}, and \textit{Heterogeneity}. The available variants are all unweighted, but allow handling undirected, incoming and outgoing edges.

\subsubsection{Graph-Focused Topological Measures}
\label{sec:GraphMeasures}				
A simple way to obtain graph-focused measures is to consider a vertex-focused measure and compute some statistic over the vertex set of the graph. This is what we do for all of the $21$ measures described in the previous section, by averaging them over the whole graph. This also holds for all the variants (weighted and/or directed) of these measures. But there are also measures defined specifically for the graph scale: like before, we distinguish them based on their scope.

\paragraph{Microscopic Measures} 
First, we use very classic statistics describing the graph size: the \textit{Vertex} and \textit{Edge Counts}. We also compute the \textit{Density}, which corresponds to the ratio of the number of existing edges to the number of edges in a complete graph containing the same number of vertices. In other words, the density corresponds to the proportion of existing edges, compared to the maximal possible number for the considered graph. In our context, these measures allow assessing the number of users considered in a context period (Vertex Count), and the general intensity of their communication during this period (Edge Count). The Density can be viewed as a normalized Edge Count that is more likely to be useful when comparing graphs of different sizes.

The \textit{Global Transitivity} (or \textit{Global Clustering Coefficient}~\cite{Luce1949} is the graph-focused counterpart of the Local Transitivity. It corresponds to the proportion of closed triads among connected ones, where a closed triad is a $3$-clique (\textit{i.e.} a triangle) and a connected triad is a subgraph of $3$ vertices containing at least $2$ edges. This proportion measures the prevalence of triadic closure in the graph. In our context, it assesses how likely two users communicating with the same person are to directly exchange messages themselves. We only have access to the undirected unweighted version of this measure.

The \textit{Reciprocity} \cite{Wasserman1994} is defined only for directed graphs. It corresponds to the proportion of bilateral edges over all pairs of vertices. In our networks, a low reciprocity would indicate that certain users do not respond to others.

The \textit{Degree Assortativity} (or \textit{Assortativity} for short) \cite{Newman2002} measures the homophily of the graph relatively to the vertex degree. The homophily is the tendency for vertices to be connected to other similar vertices (in this case: of similar degree). It is based on the correlation between the series constituted of all pairs of connected vertices. We use both directed and undirected variants of this measure. In our conversational networks, this measure could help detect situations where users do not participate to the conversation at the same level.

\paragraph{Macroscopic Measures}
A number of macroscopic measures are connectivity-based. The \textit{Weak Component Count} corresponds to the number of maximally connected subgraphs. In such a subgraph, there is a path to connect any pair of vertices. For our conversational networks, this could correspond to a conversation, whose participant do not necessarily talk directly to each other. However, due to the use of a sliding window, we expect our graphs to be connected (\textit{i.e.} only one weak component), even if by very weak edges. In this case, a conversation is more likely to correspond to other substructures based on more relaxed definitions, such as cliques or communities. For directed graphs, we also consider the \textit{Strong Component Count}: a strong component is similar to a weak one, except it is based on \textit{directed} paths. We suppose that, in our networks, we are more likely to get several \textit{strong} components, since users do not necessarily exchange in a bilateral way.

The \textit{Cohesion} (or \textit{Vertex Connectivity}) of a graph corresponds to the minimal number of vertices one needs to remove in order to make the graph disconnected (\textit{i.e.} have several components)~\cite{White2001}. The \textit{Adhesion} (or \textit{Edge Connectivity}) is similar, but for edges. In our conversational networks, these measures can be related to the level of participation to the considered conversation: the higher their values, and the higher this level. But high values can also denote the presence of several distinct conversations in the context period. Both measures are defined for directed networks.

As mentioned before when describing the nodal measures, we check whether the targeted user is an articulation point. We also compute the \textit{Articulation Point Count}, \textit{i.e.} the total number of articulation points in the graph. This measure is related to the Cohesion, since there are no articulation point if the Cohesion is larger than $1$. The implementation we use handles only undirected graphs. In our context, the number of articulation points could be related to the presence of several conversations (articulation points corresponding to gateway users between them). It could also reflect situations where a conversation lasts a very long time, and some groups of users loose interest and get disconnected from the active users. Another possibility is the occurrence of a flood-type situation: a user sends a flurry of messages into the channel to kill a conversation, then leave, and a different group of users later takes possession of the channel to start its own conversation.

We also use three distance-related measures. The first is the \textit{Diameter}, which corresponds to the largest distance found in the graph, \textit{i.e.} the length of the longest shortest path. It also corresponds to the largest Eccentricity over all vertices. We use (un)weighted and (un)directed variants. The second is the radius, which is the smallest Eccentricity over all vertices. We use its undirected, incoming and outgoing variants. The third is the \textit{Average Distance}, which is the average length of the shortest paths processed over all pairs of vertices. We use its unweighted (un)directed variants. In our networks, the distance is related to the separation between users, in terms of interaction. A large Diameter means that a user can be many intermediaries away from exchanging directly with another user. This could be caused, for instance, by the occurrence of several distinct conversations in the considered context period, or by a very long conversation loosing and gaining users through time. This observation also holds for the Radius and Average Distance, which provide a slightly different perspective on the same aspect of the graph.

\paragraph{Mesoscopic Measures} 

We process the total \textit{Clique Count} in the network, where a clique is a complete induced subgraph. As mentioned before, this can be related to the number of conversations occurring in the context period, or to number of subgroups of users participating in the same conversation.

Like before with vertex-focused measures, we use the InfoMap algorithm to detect the community structure~\cite{Rosvall2008}. Based on this partition, we compute two measures: the \textit{Community Count} and the \textit{Modularity}~\cite{Newman2004e}. The latter assesses the quality of the detected community structure, \textit{i.e.} how internally cohesive and externally disconnected the communities are. We use both weighted and unweighted variants of the Modularity.

\section{Experiments}
\label{sec:exp}

This section describes our experimental setup and results regarding the automatic detection of abusive messages in chat logs. In Section~\ref{sec:exp-setup}, we present our dataset and the general architecture of our classification system. Because we expect some of our features to be redundant, we conduct a correlation study of our feature set in Section~\ref{sec:exp-features-dep}. We present general results and the effect of our various graph extraction parameters in Section~\ref{sec:res-build}. In Section~\ref{sec:res-three-graphs}, we investigate the temporal aspects of the system --specifically what happens when we train our models based on the features extracted from only one of the three graphs (\textit{Before}, \textit{After} of \textit{Full}), or some of their combinations. We then examine the importance of weight and directionality in Section~\ref{sec:res-weigths-direction}, before investigating the potential for computational optimization through feature selection in Section~\ref{sec:res-abla}. Finally, in Section~\ref{sec:res-baseline}, we compare the performance obtained using the best configuration of our framework with the selected baselines.

\subsection{Experimental Setup}
\label{sec:exp-setup}
We have access to a database of $4,029,343$ messages that were exchanged by the users of the browser-based multi-player game \textit{SpaceOrigin}, a French-language Massively Multiplayer Online Game. In this database, $779$ messages have been flagged by one or more users as being abusive, and subsequently confirmed as abusive by the human game moderators: they constitute our \textit{Abuse} class. Each message belongs to a unique communication channel. A total of $226$ distinct users have authored these abusive messages. We further extract $2,000$ messages at random from the messages not confirmed as abusive, to constitute the \textit{Non-abuse} class. Note that all the results we discuss in this article are relative to the \textit{Abuse} class. 

\begin{figure}[!t]
    \centering
	\includegraphics[width=0.99\linewidth]{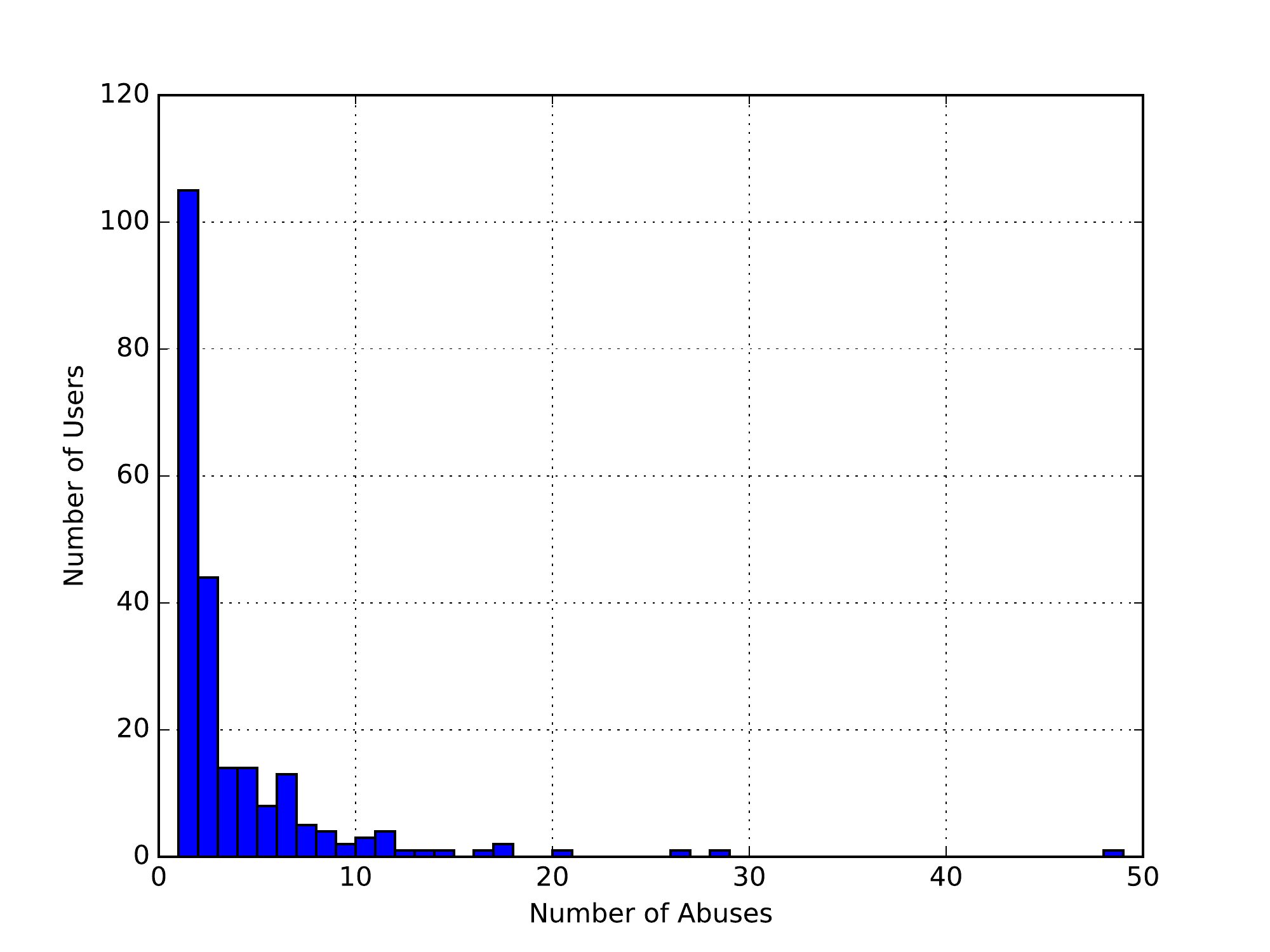}
    
    \caption{Distribution of the number of abuse cases by user.}
	\label{fig:abuse_distro}
\end{figure}

We previously experimented with this dataset in~\cite{Papegnies2017a, Papegnies2017}. However, since then we have detected certain inconsistencies in the database, preventing us from retrieving the context of certain messages. We cannot apply our classification method to them, so we discard them for the work presented here. Note that this concerns both classes. Moreover, our tests show that removing those samples does not significantly impact our previous performances. The resulting dataset is constituted of $1,890$ messages in the \textit{Non-Abuse} class and $655$ messages in the \textit{Abuse} class. Figure~\ref{fig:abuse_distro} shows the distribution of abuse cases by user. It suggests that most abusive users need only a few warnings before mending their ways, but it also shows that some users are exceptional in the number of abuses they commit.

Because of the relatively small dataset, we set up our experiment for a $10$-fold cross-validation. We split the dataset into $10$ same-sized parts containing the same ratio of abusive to non-abusive messages. We use a 70\%-train / 30\%-test split, which means, for each run of the cross-validation, the train set is composed of $7$ of those parts while the test set is composed of the remaining $3$. We use Python-iGraph~\cite{Csardi2006} to extract the conversational networks and process the graph-based features for each message. As a classifier, we use an SVM (Support Vector Machine), implemented in the \textit{Sklearn} toolkit~\cite{Pedregosa2011} under the name SVC (C-Support Vector Classification).

We mainly experiment with $4$ different sets of features: \textit{Full}, \textit{Before}, \textit{After} and \textit{All}. For a given message, \textit{Before}, \textit{After} and \textit{Full} correspond to all the topological measures computed for the \textit{Before}, \textit{After} and \textit{Full} graphs, respectively. \textit{All} is the union of all three sets, \textit{i.e.} it includes all topological measures for all $3$ graphs.

In the remainder of this section, we occasionally provide computational time requirements. For context, the times that we provide correspond to single-threaded calculations performed on an Intel Xeon CPU E5-2620 v3s, clock speed 2.5 GHz and 15 MB cache.

\subsection{Feature Dependence Study}
\label{sec:exp-features-dep}
Each considered topological measure was originally defined to characterize a graph in a specific, distinct way. So, in theory, they could all be independent for a given graph, and thus all necessary to describe it completely. But in practice, according to the structure of the considered graph, some of them can be statistically dependent, and therefore redundant. In order to get a better understanding of the way these topological measures behave on our conversational graphs, we compare all the computed features using Pearson's correlation coefficient. In our context, where these features are later fetched to a classifier, only the strength of the association is relevant, \textit{i.e.} the absolute value of the correlation (not its sign). We identify clusters of highly correlated features using the \texttt{hclust} function of the \texttt{R} language, which implements a standard hierarchical cluster analysis method, with average linkage. We use the \textit{Silhouette} measure~\cite{Rousseeuw1987} as a criterion to select the best cut in the produced dendrograms. To keep the description short, we only focus on the most interesting results. 

A very small number of features are constant over all instances of the corpus, which means they have no discriminative power at all. For all three types of networks, the number of weak components is always $1$, which means they are always (weakly) connected. This can be explained by our use of a sliding window: even if the context period contains two separate conversations, they will be connected, possibly by a edge of quasi-zero weight. In the \textit{After} and \textit{Full} graphs, the targeted user is never an \textit{articulation point}. Moreover, in the \textit{Full} graph, the number of articulation points is always zero. We already know that the graphs are connected, so this zero value means no single vertex removal can disconnect them. 

A few features are quasi-independent, in the sense they display almost no correlation with any other feature. This is the case of certain variants of the Power, Subgraph and Alpha centralities. From this point of view, they differ from the other spectral measures, which are overall strongly correlated. Certain variants of measures focusing on connectivity (Strong Component Count, Adhesion, Cohesion and Radius) are also independent. And it is the case for a number of mesoscale features too, all of them based on the community structure. The fact that these features are only weakly (if at all) correlated to the others makes them singular, in the sense they are the only ones to capture certain structural changes in the conversational graphs. But it does not imply they have any particular discriminative power regarding the classification task at hand. However, they must be closely monitored in the rest of our experiments, because they constitute good candidates.

The rest of the features form highly correlated clusters. As explained in Section~\ref{sec:methods-features}, for each topological measure we consider several variants to define our features. Those can be direction- (undirected, incoming and outgoing variants of the same measure), weight- (unweighted vs. weighted) and average-based (nodal measure processed for the targeted vertex vs. averaged over the whole graph), and of course these different traits can be combined. Our correlation study shows that certain variants of the same measure are strongly related, \textit{i.e.} are placed in the same cluster, which makes them redundant for our purpose. A very large number of measures, mainly distance- and community-based, are direction-independent, \textit{i.e.} all their direction-based variants are strongly correlated. This indicates that most of the time, considering the direction of the interactions between users does not bring any additional information. This effect is clearly much less marked for average-based and weight-based variants. Thus, unlike direction, weight seems like an important aspect of our graphs, and averaging measures over all vertices also seems to bring some relevant information. This opens an interesting perspective, as it may be promising to explore other statistics such as the standard deviation. 


The strong correlations between variants of the same measure form small tight clusters, but there are also larger clusters gathering a number of variants based on distinct measures. They tend to concern measures relying on a similar concept, \textit{e.g.} community structure. But there are also more surprising associations, for instance for each type of graph, the larger cluster contains $20$--$30$ features related to Authority/Hub Scores, Eigenvector centrality and PageRank (which are all spectral measures), but also Eccentricity, Diameter, Radius and average Distance (all distance-based). 

Overall, we observe different behaviors, which cannot be explained only by the various characteristics of the features (micro/meso/macroscopic, un/directed, un/weighted, \textit{Before}/\textit{After}/\textit{Full} graphs). This supports our decision to adopt an exploratory approach to identify the most appropriate features for our classification problem. The detected clusters of correlated features will be useful later to ease the interpretation of the classification results, as features belonging to the same cluster can be considered as interchangeable. It is worth noticing that these clusters are based on a \textit{linear} definition of the correlation. But the classifier we use is able to detect non-linear associations, and may therefore consider that some of these clusters are equivalently informative for the problem at hand (by detecting additional or stronger relations between the features).


\subsection{Impact Of Graph Extraction Parameters}
\label{sec:res-build}
As explained in Section~\ref{sec:methods-netext}, our graph extraction method has three important parameters: 1) size of the context period; 2) size of the sliding window, and 3) weight assignment strategy. In this section, we explore how the classification performance varies depending on these parameters. Our goal here is both to get a better understanding of the parameters role, and to identify the most appropriate values without having to use brute force.

\begin{figure}[t]
	\centering
	\includegraphics[width=0.99\linewidth]{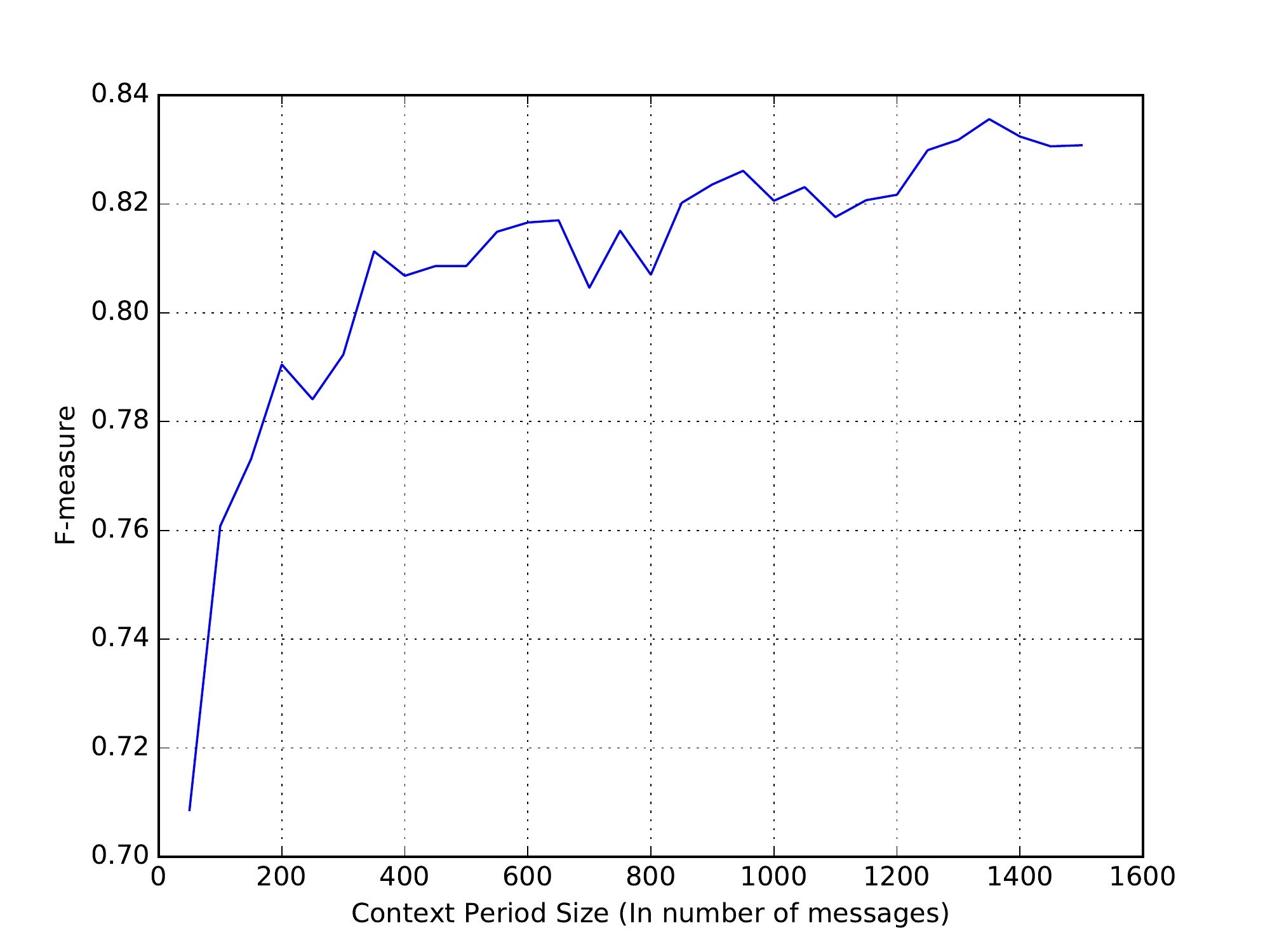}
    
    \caption{Classification performance ($F$-measure) as a function of the context period size, for a sliding window of $10$ messages and using the \textit{Recursive} weight assignment strategy ($f_R$).}
	\label{fig:res_context_size}
\end{figure}

As a reminder, the context period is the sequence of messages considered to classify the targeted message, and symmetrically built around this message. We expect it to have a strong effect on the classification performance, depending on its size. If it is too small, one can suppose it only includes a part of the conversation containing the targeted message, and therefore lacks some information necessary to make a proper decision regarding the abusive nature of this message. On the contrary, if it is too large, we assume it contains several conversations having nothing to do with the targeted message, which should also result in lower classification performance. In summary, when using a growing context period, we expect the classification performance to increase, then reach a plateau corresponding more or less to the typical duration of a conversation, and then decrease as the context period contains more and more noise (\textit{i.e.} information not related to the targeted message).

Figure~\ref{fig:res_context_size} shows the evolution of the classification performance, expressed in terms of $F$-measure for the \textit{Abuse} class, as a function of the context period size, as for it expressed in numbers of messages. We fix the sliding window to $10$ messages, and the recursive strategy $f_R$ to assign weights. All available features (\textit{All} feature set) are used during the classification. We choose these extraction parameters because earlier testing showed that the recursive strategy yielded the best performance, and this sliding window size provides a good trade-off between a graph that would be very sparse and therefore not informative enough, and one that would be very dense and thus too noisy.

It appears that our assumption is only partially verified: the performance first increases with the context period size. However, it does not reach a plateau as expected, and, on the contrary, seems to go on increasing, albeit more and more slowly, as if it was logarithmically depending on the context period size. The maximal performance is obtained for a size of $1,350$ messages, but it is possible that even higher values can be obtained for larger context periods (which we did not check due to computational limitations). This means that our assumption regarding that large context periods would bring mainly additional noise is incorrect, because, on the contrary, they convey more relevant information concerning the classification task.

\begin{figure}[t]
	\centering
	\includegraphics[width=0.99\linewidth]{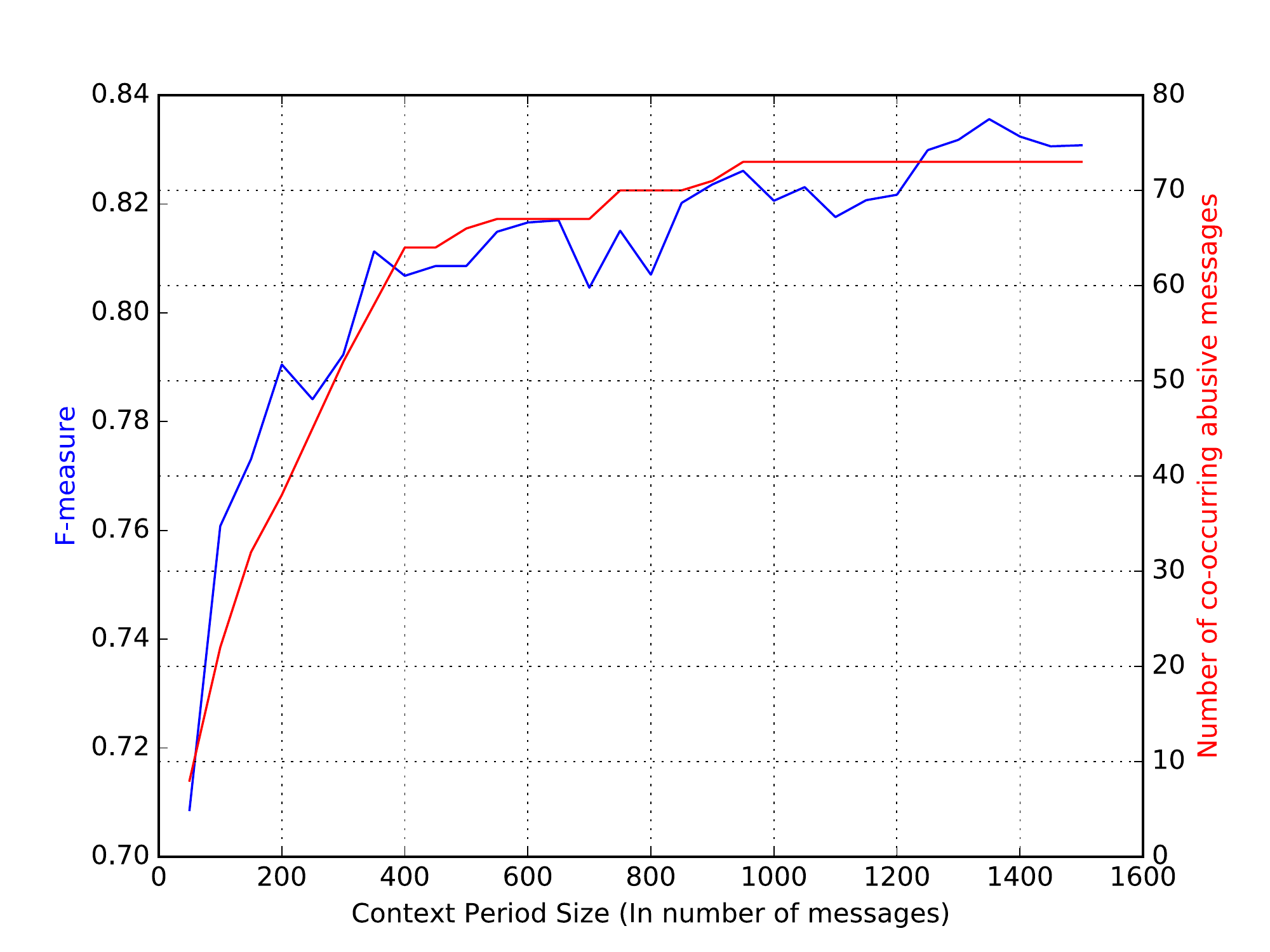}
    
    \caption{Classification performance (blue) and number of abuse occurrences in the context period (red), as functions of the context period size.}
	\label{fig:collisions}
\end{figure}

We manually investigate a sample of our dataset to understand this trend. From reading a number of conversation logs, it first appears that conversation boundaries are not well defined, and so there is no typical duration for a conversation: this can explain the absence of a plateau in the plot. Furthermore, we based our assumptions with regard to performance on the idea that an abusive message has a specific impact on what happens after it is posted. Specifically, the conversation would show markers of normality before the message occurs and quickly devolves after that. As it turns out, in conversations where an abusive message is found, the author of the abusive message usually has been around for a while, and the message that is actually flagged and confirmed as abusive is not his first suspicious message. We assume that the classifier can take advantage of this type of situations, therefore invalidating our previous assumption that a large context period would only bring noise. 

Note that more than one message of the \textit{Abuse} class can co-occur, \textit{i.e.} can appear in the same context period, if it is large enough, which also supports our point. This is generally due to a single user sending multiple abusive messages in quick succession, or because the conversation devolves into name-calling following an initial abusive message. Figure~\ref{fig:collisions} shows the number of co-occurring abusive messages, as well as the $F$-measure performance, as functions of the size of the context period. There is a very strong match between both series, even if not perfect. This seems to back our assumption of the classifier taking advantage of the potential abuse cases happening around the targeted message. All these observations regarding the co-occurrence of abuse expose a couple of interesting perspectives: 1) user models can presumably yield features useful for classification, and 2) a text-based model of the whole conversations would also likely be useful.

\begin{figure}[t]
    \centering
	\includegraphics[width=0.99\linewidth]{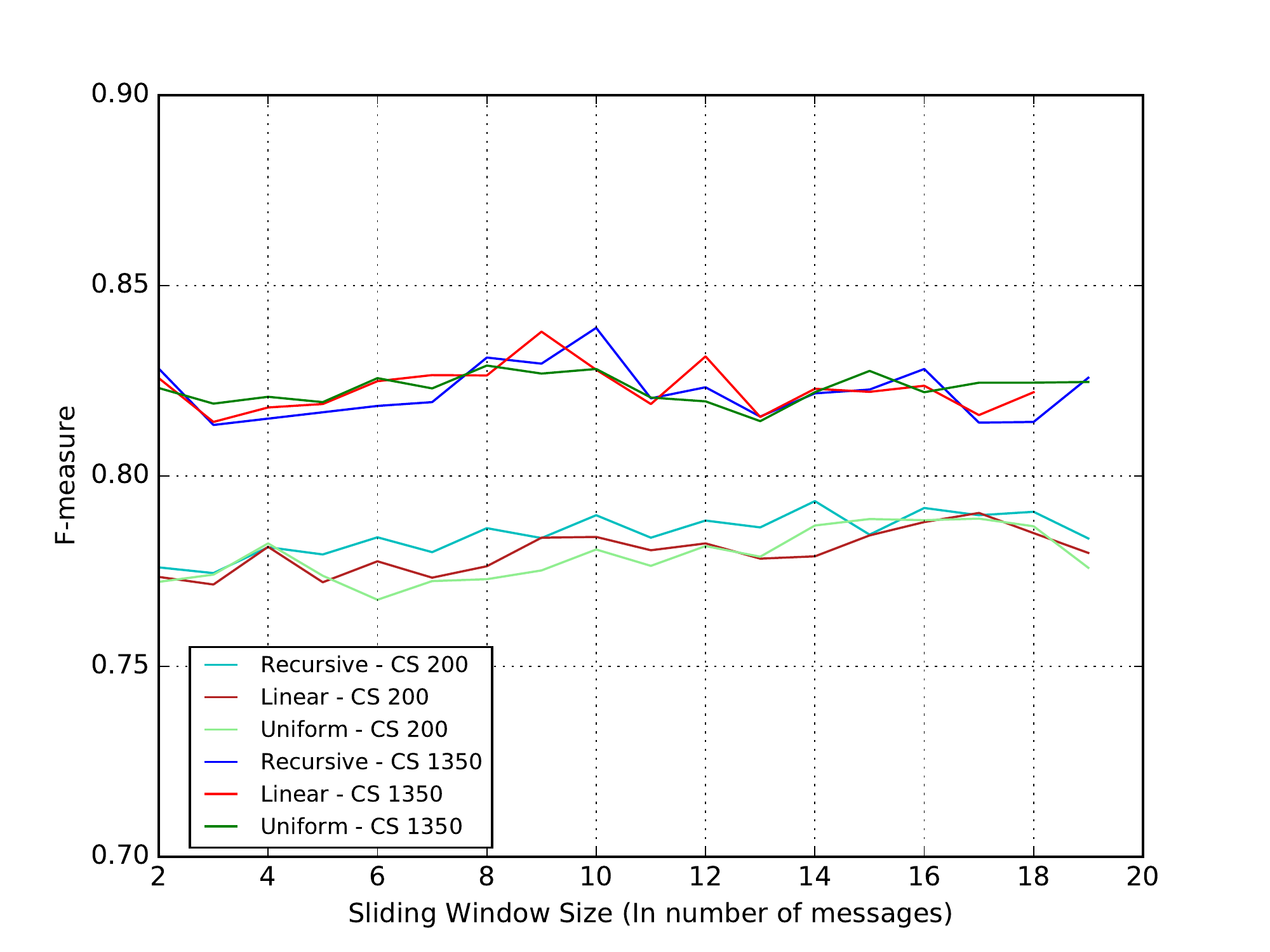}
    
    \caption{Classification performance ($F$-measure) for the $3$ considered weight assignment strategies (\textit{Uniform}, \textit{Linear} and \textit{Recursive}) and $2$ context period sizes ($200$ and $1,350$ messages), as a function of the sliding window size.}
	\label{fig:res_window_size}
\end{figure}

We now explore the impact of the window size and weight assignment strategy on the overall performance. Figure~\ref{fig:res_window_size} shows the evolution of our performances for two fixed sizes of context periods ($200$ and our previously obtained optimum of $1,350$). The maximal window size considered is $19$, which corresponds to almost twice the default GUI limitation (cf. Section~\ref{sec:net-slide}).

For our optimal context size, we obtain the best results for a window size of $9$ and the Linear assignment strategy, and for the window size $10$ and the Recursive assignment strategy. Both of those strategies give greater importance to temporal proximity. Overall, there is no much difference between our weight assignment strategies. It seems that the specific values of the weights are not as important as their relative ordering. It is also worth noting that those window sizes are very close to the natural limitation of the GUI, which means they likely best capture the intended recipients of any given message.

\subsection{Temporal Aspects}
\label{sec:res-three-graphs}
The results shown until now are all obtained using the \textit{All} feature set, \textit{i.e.} the features resulting from the calculation of all topological measure variants for all three individual graphs (\textit{Full}, \textit{Before} and \textit{After}). However, using the \textit{Full} or \textit{After} graphs restricts the possible use cases for the system to tasks that do not require taking a decision as soon as the message to classify becomes available since the ``future'' context of the targeted message is taken into account. In order to have a system that is capable of doing so, we must investigate the impact of the \textit{Before} features. Studying the features from the three individual graphs also allows us to get a better understanding of the system by providing a qualitative and accurate analysis of each part of the context.

\begin{figure}[!t]
    \centering
	\includegraphics[width=0.99\linewidth]{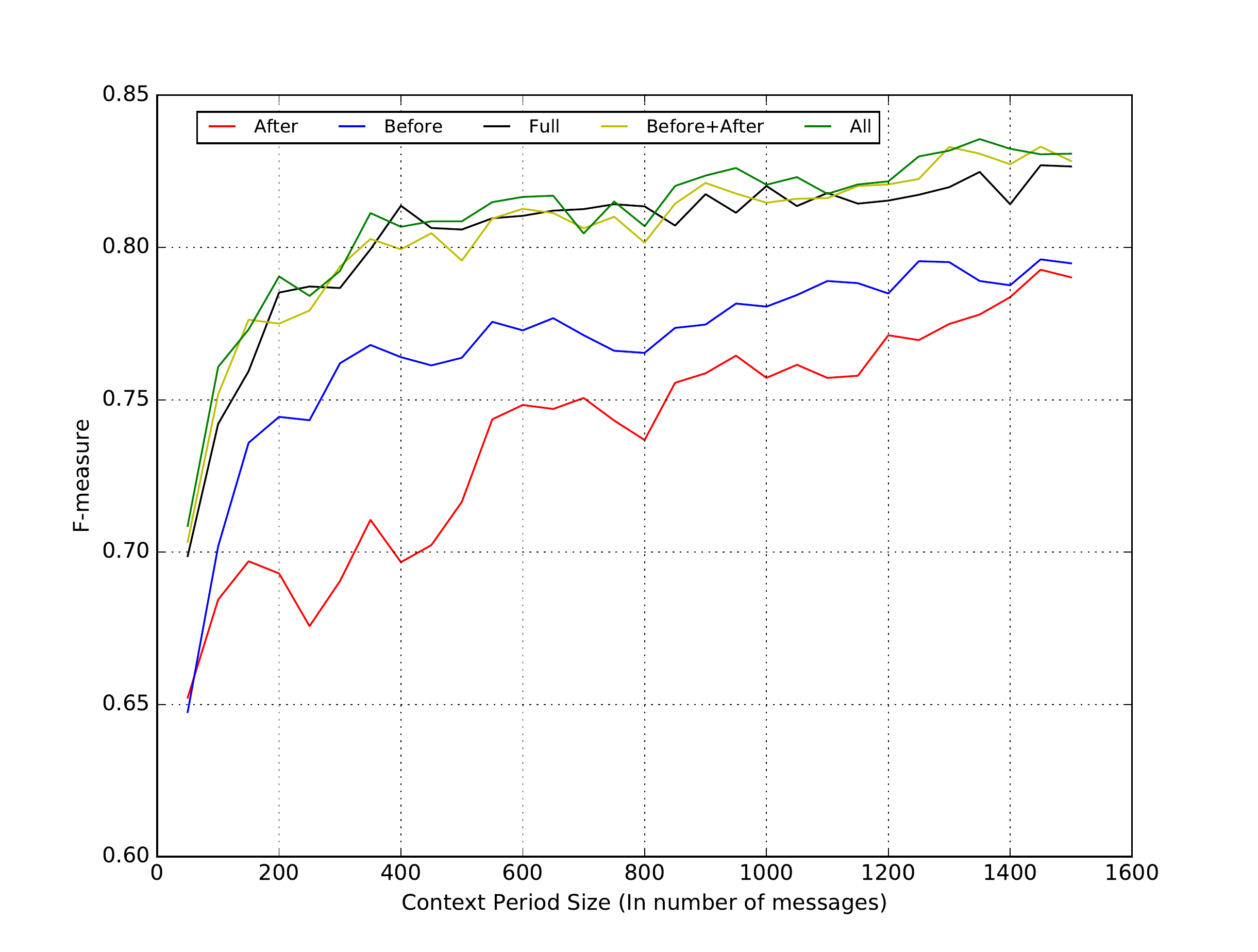}
    
    \caption{Classification performance ($F$-measure) as a function of the context period size (in messages), for the $4$ considered feature sets (\textit{Before}, \textit{After}, \textit{Full}, \textit{All}) as well as a combination of the first two of them (\textit{Before+After}). }
	\label{fig:res_full_vs_before_vs_after_vs_all}
\end{figure}

Figure~\ref{fig:res_full_vs_before_vs_after_vs_all} shows the results obtained for classifiers built using combinations of the available feature sets: \textit{After}, \textit{Before} and \textit{Full} correspond to each graph considered separately, whereas \textit{Before+After} denotes the union of the \textit{Before} and \textit{After} sets, and \textit{All} represents the set of all computed features (\textit{Before}, \textit{After} and \textit{Full}). The conversational graphs used for these experiments are extracted using the \textit{Recursive} assignment strategy $f_R$ and a sliding window of $10$ messages. Unless stated otherwise, we use these parameters in the rest of this paper as they match the best performance obtained during our greedy search of the parameter space (Section~\ref{sec:res-build}).

Since the main idea behind our approach is to detect the nature of a message based on the reaction it triggers in the community, it is not surprising to see that the \textit{After} feature set (in red) reaches an acceptable performance level on this task. However, what is surprising is that by using only the \textit{Before} feature set (in blue), the system performs much better on the same task (at least for small context periods). This suggests that the interactions occurring before the targeted message reveal more about its abusive nature than those happening after. 

Nevertheless, when using large context periods, the performances obtained for \textit{Before} and \textit{After} get very similar. This indicates that what is important is not whether the messages used to extract the conversational graph \textit{precede or follow} the targeted message, but rather \textit{how many} of these messages are used. This supports our previous finding, regarding the fact that the context period size is the most important parameter of our graph extraction procedure. Moreover, based on the same observation, one could also think that the \textit{Before} and \textit{After} graphs convey approximately the same information, when considering large context periods, because the corresponding performances are roughly the same. However, this is disproved by the results obtained for the union of the \textit{Before} and \textit{After} feature sets (in yellow): the classification performance is noticeably higher, which means both types of graphs do not completely overlap, informationally speaking.

It is worth noticing that we get almost the same performance with the \textit{Full} feature set as with \textit{Before+After}. One could assume that the use of two distinct graphs built on either sides of the targeted message would help better characterizing it, compared to the \textit{Full} feature set, which covers the same time span based on a single graph. Indeed, when extracting the latter, the sliding window passes through the targeted message, and is likely to smooth the potentially relevant topological changes occurring right around it. However, even if the performance gap between \textit{Before} and \textit{After} seems to widen when the context period gets larger, the difference is not clear, so this assumption is not verified. This means that the single \textit{Full} graph is as approximately as informative as the joint use of both \textit{Before} and \textit{After} graphs. The latter option procures more flexibility in the possible application scenarios, but it contains twice as many features, and therefore requires roughly twice the computational time.

This observation is confirmed when we consider the \textit{All} feature set (in green), which contains all features for all $3$ graphs. As expected, it is the best performing feature set overall, since it is the union of all the other considered feature sets. However, the obtained results are only marginally better than for \textit{Full} and \textit{Before+After}. This means that the information conveyed by the \textit{Full} and \textit{Before+After} feature sets essentially overlaps: using their union does not bring any noticeable performance increase. 

\subsection{Impact of Weights and Directions}
\label{sec:res-weigths-direction}
We now investigate how considering the edge weights and directions in our features affects the classification performance. Based on the \textit{All} feature set, we define $4$ new feature sets, characterized by their focus on unweighted undirected (UU), unweighted directed (UD), weighted undirected (WU), and weighted directed (WD) measures, respectively. Concretely, each set includes the same group of core features, which are conceptually not concerned by the notion of weight or direction. This core is completed by features designed to consider or ignore weights or directions. For instance, the Clique Count is a core feature, whereas each one of the $4$ variants of the Diameter appears in a specific set.

\begin{figure}[!t]
    \centering
	\includegraphics[width=0.99\linewidth]{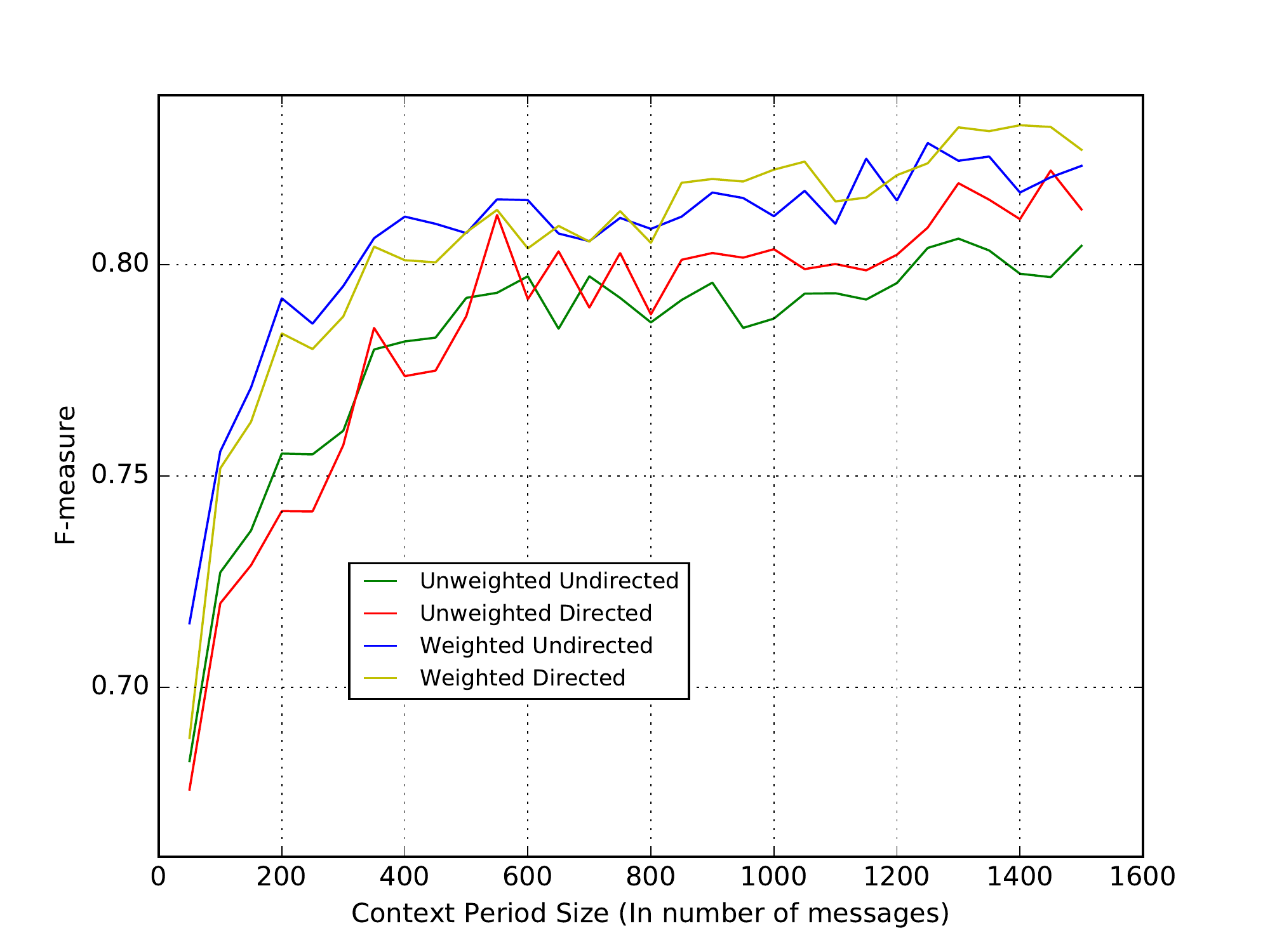}
    
    \caption{Performance ($F$-measure) obtained for the (un)directed and (un)weighted feature sets, as a function of the context period size.}
	\label{fig:res_uu_ud_wu_wd}
\end{figure}

Figure~\ref{fig:res_uu_ud_wu_wd} displays how the corresponding classification performance (in terms of $F$-measure) evolves as a function of the context period size. It appears that both weighted feature sets (blue and yellow) dominate their unweighted counterparts (green and red) over the considered interval. This seems to confirm our assumption from Section~\ref{sec:methods-netext}, regarding the fact that weights can help discriminate between certain structures of conversations, and/or distinguish consecutive conversations. 

There is a similar effect for directions, but it is much weaker, as each directed feature set (red and yellow) only partially dominates its undirected counterpart (green and blue). This is consistent with our observation from Section~\ref{sec:exp-features-dep}, regarding the high correlation noticed for certain measures, between their undirected and directed variants (and therefore their potential lack of discriminative power). This indicates that the direction of edges is not as relevant as their weight relatively to the classification task at hands. Yet, the best performance is reached when using both weights and directions. If the additional computational cost is not too high (and it is generally not the case), it is therefore worth using directed features.

\subsection{Feature Contributions}
\label{sec:res-abla}
In order to estimate the discriminative power of our features with regard to this classification task, we use a recursive feature elimination method. It takes a given feature set as input, and outputs its subset of so-called \textit{Top Features} (TF). These are the minimal subset of features allowing to reach 97\% of the performance obtained when considering the input feature set. In order to identify these top features, we apply an iterative method based on \textit{Sklearn}. This toolkit allows us to fit a linear kernel SVM with the values of the input feature set, and provides a ranking of the individual features in that set, reflecting their relevance to the classification task. We then drop the least important feature, and train a new model using all the remaining features. We repeat this process until the classification performance reaches the targeted minimal threshold of 97\% of the original $F$-measure score. 

\begin{table}[!ht]
\centering
\caption{Comparison of the performances obtained with the feature sets (\textit{All}, \textit{Before}, \textit{Full}, \textit{After}) and their subsets of \textit{Top Features} (TF). The total runtime is expressed as \textit{day}:\textit{h}:\textit{min}:\textit{s}.}
  \begin{tabular} { |l|r|r|r|r| }
    \hline
    \textbf{Feature} & \textbf{Number of} & \textbf{Total} & \textbf{Average} & \textbf{$F$-} \\
    \textbf{set} & \textbf{features} & \textbf{Runtime} & \textbf{Runtime} & \textbf{measure} \\ 
    \hline
    \textbf{All} & 459 & 4:15:29:24 & 157.71 s & \textbf{83.89} \\
    \textbf{All-TF} & 10 & 53.92 & 0.02 s & \textbf{82.65} \\ 
    \hline
    \textbf{Before} & 153 & 1:05:51:06 & 42.23 s & 79.03 \\
    \textbf{Before-TF} & 8 & 29.68 & 0.01 s & 79.02 \\ 
    \hline
    \textbf{After} & 153 & 1:06:04:15 & 42.54 s & 78.28 \\
    \textbf{After-TF} & 11 & 1:30.72 & 0.04 s & 76.01 \\ 
    \hline
    \textbf{Full} & 153 & 2:03:34:02 & 72.94 s & 82.17 \\
    \textbf{Full-TF} & 6 & 2:38.37 & 0.06 s & 82.65 \\ 
    \hline
  \end{tabular}
  \label{tab:SubsetPerfs}
\end{table}



We first apply this recursive feature elimination process to the \textit{All} feature set in order to identify the overall best features, then do the same with the \textit{Before}, \textit{After}, \textit{Full} feature sets, for comparison purposes. Table~\ref{tab:SubsetPerfs} presents the performances and computational time costs measured for each of these complete feature sets, as well as for their respective \textit{Top Feature} subsets (TF). It appears that, for all $4$ feature sets, using the top features during the classification allows reducing the runtime by up to $4$ orders of magnitude, while retaining at least 97\% of the $F$-measure value, which is very interesting from an application perspective. It means that the longer features to process do not bring more discriminative power than the shorter ones, regarding the classification task at hand (at least for our dataset). The fourth column describes the average runtime \textit{by message}, and shows that the classifier could operate in real-time when limited to the top features.

It is important to notice that feature computation is by far the most computationally expensive step of our framework. By comparison, extracting the conversational graphs for the full corpus takes around $3$ minutes, and performing the whole cross-validation (\textit{i.e.} 10-fold training and testing) only $1$ minute. The time required to compute our features depends on the size of the conversational graph, in terms of number of vertices and/or edges. The graph size is affected, in turn, by the number of users involved in the conversation (number of vertices) and the density of exchanged messages (number and weights of the edges). These characteristic are bounded by social and ergonomic (\textit{e.g.} user interface) constraints, and therefore be assumed as independent from the corpus size. The scalability of our method therefore depends on that of the tool selected to perform the classification step: it is an SVM in this work, but the end-user is free to use any other classifier instead.

As explained in Section~\ref{sec:exp-features-dep}, certain of our features are strongly correlated, which led us to identify clusters of interchangeable features. Studying the top features would result in missing this information: we must consider their \textit{clusters} instead. Table~\ref{tab:SubsetAll} displays the $9$ clusters corresponding to the top features obtained for the \textit{All} feature set. For matters of space, we discuss in detail this sole feature set only. Note that these clusters generally contain several variants of one (or more) topological measure(s), as indicated by the $4$ last columns. For instance, Cluster 10 contains all variants of average weighted and unweighted undirected Eigenvector Centrality for all $3$ types of graphs (\textit{Before}, \textit{After}, \textit{Full}). Also note that a letter \textit{G} in the \textit{Scale} column can either refer to a naturally graph-scale feature, or to a vertex-scale feature averaged over $V$ (cf. Section~\ref{sec:methods-features}). For completeness, the proper top features are represented in bold.

\begin{table}[!ht]
	\centering
    \caption{Clusters containing the top features (in bold) obtained for the \textit{All} feature set. The letters in the \textit{Graph} column stand for \textit{Before} (B), \textit{After} (A) and \textit{Full} (F). Those in the \textit{Scale} column mean \textit{Graph-scale} (G) or \textit{Vertex-scale} (N). Those in the \textit{Wght.} and \textit{Dir.} columns have the same meaning as in Table~\ref{tab:FeatureList}.}
    \begin{tabular}{|r|l|l|l|l|l|}
        \hline
     	\textbf{Clust.} & \textbf{Measure} & \textbf{Graph} & \textbf{Wght.} & \textbf{Dir.} & \textbf{Scale} \\
        \hline
        \multirow{4}{*}{9} & Clique Count & A/F & -- & -- & G \\
		 & Burt's Constraint & A/F & W & -- & G \\
		 & \textbf{Coreness Score} & A/\textbf{F} & -- & U/\textbf{I}/O & \textbf{G} \\
		 & Degree Centrality & A/F & U & U/I/O & G \\
		 & Strength Centrality & A/F & W & U/I/O & G \\
        \hline
		\multirow{11}{*}{10} & Assortativity & A/B/F & -- & U/D & G \\
		 & Density & A & -- & -- & G \\
		 & Diameter & A & U & U/D & G \\
		 & Average Distance & A & U & U/D & G \\
		 & Radius & A & U & U & G \\
		 & Hub/Authority Scores& A/F & W & D & G \\
		 & Burt's Constraint & A/F & U & -- & G/N \\
		 & Closeness Centrality & A & U & U/I/O & G \\
		 & Eccentricity & A & U & U/I/O & G \\
		 & Eigenvector Centrality & A/B/F & U/W & U & G \\
		 & \textbf{PageRank Centrality} & \textbf{A}/F & \textbf{U}/W & U/\textbf{D} & G/\textbf{N} \\
        \hline
		\multirow{2}{*}{41} & Degree Centrality & A/F & U & U/I/O & N \\
		 & \textbf{Strength Centrality} & A/\textbf{F} & \textbf{W} & U/I/\textbf{O} & \textbf{N} \\
        \hline
		\multirow{2}{*}{49} & \textbf{Vertex Count} & A/\textbf{F} & -- & -- & G \\
		 & Betweenness Centrality & A/F & U/W & U/D & G \\
        \hline
		\multirow{2}{*}{110} & Density & B & -- & -- & G \\
		 & \textbf{Closeness Centrality} & \textbf{B} & \textbf{W} & U/I/\textbf{O} & \textbf{G}/\textbf{N} \\
        \hline
		\multirow{3}{*}{118} & Average Distance & B & U & U/D & G \\
		 & Hub/\textbf{Authority Scores} & \textbf{B} & \textbf{W} & \textbf{D} & \textbf{G} \\
		 & PageRank Centrality & B & U/W & U/D & G/N \\
        \hline
		119 & \textbf{Hub}/Authority Scores & A/\textbf{B} & \textbf{U} & \textbf{D} & \textbf{N} \\
        \hline
		172 & \textbf{Reciprocity} & \textbf{A} & -- & \textbf{D} & G \\
        \hline
		177 & \textbf{Closeness Centrality} & \textbf{A} & \textbf{W} & \textbf{U}/O & G/\textbf{N} \\
        \hline
  	\end{tabular}
	\label{tab:SubsetAll}
\end{table}

Cluster 9 contains only micro- (Degree, Strength, Burt's Constraint) and meso-scopic (Clique Count, Coreness) features describing the \textit{After} and \textit{Full} graphs. Moreover, all of them are graph-scale (as the vertex-scale measures are averaged over the graph). In contrast, Cluster 41 focuses on the same graphs and also contains the Degree and Strength, but as vertex-scale features this time. Put differently, Cluster 41 can be viewed as a vertex-scale counterpart of Cluster 9. This indicates that the microscopic characteristics of both the targeted vertex and the whole graph are relevant to our classification task.

Cluster 10 is very large and contains almost only graph-scale features. It focuses mainly on distance-based (Diameter, Average Distance, Radius, Closeness, Eccentricity) and spectral (Hub/Authority, Eigenvector, PageRank) macroscopic measures. Like the previous clusters, it essentially contains features computed on the \textit{After} graph, but unlike them, it includes only a few features from the \textit{Full} graph. Nevertheless, it appears as quite complementary of Cluster 9, in the sense it can be considered as its macroscopic counterpart. 

Cluster 49 suggests that the Betweenness of the \textit{After} and \textit{Full} graphs mechanically increases with their number of vertices. But more importantly, it identifies the Vertex Count, \textit{i.e.} the size of the conversation after the targeted message, as one of its most discriminative aspects, relatively to our classification task. The interpretation of Clusters 172 and 177 is even clearer, as each focus on a single measure (Reciprocity and Closeness), uniquely for the \textit{After} graph. The bilateral nature of the exchanges after the targeted message, as well as how direct these are, can therefore also be considered as very important for the classification.

Clusters 110 and 118 deal only with the \textit{Before} graph. Cluster 110 includes variants of the weighted Closeness (both for the targeted vertex and in average) and Density. It can be considered as the \textit{Before} counterpart of Cluster 177, which also focuses on the weighted Closeness but for the \textit{After} graph. Cluster 118 contains distance-based and spectral macroscopic measures, mainly describing the whole graph. Thus, although much smaller, it can be seen as the \textit{Before} counterpart of Cluster 10, semantically speaking. 

Finally, Cluster 119 contains the unweighted Hub and Authority Scores of the targeted vertex, for both \textit{After} and \textit{Before} graphs. It can be opposed to both Clusters 10 and 118, which also contain Hub and Authority for the \textit{After} and \textit{Before} graphs, respectively, but in their weighted and averaged versions.

%
%
%
%
%
%

Let us summarize our observations. A number of composite clusters describe the \textit{After}/\textit{Full} (Clusters 9, 10 and 41) and \textit{Before} (Cluster 118) graphs at various scales and scopes. Two clusters focus more precisely on the Closeness, for the \textit{Before} (Cluster 110) and \textit{After} (Cluster 177) graphs. We assume that the classifier is able to take advantage of this to compare various aspects of the graphs, be it in terms of scale (Clusters 9 vs. 41), scope (Clusters 9 vs. 10) or time (Cluster 10 vs. 118, and 177 vs. 110). This means that 1) temporal aspects are useful for this classification task and 2) an abuse case is reflected by its impact on both the position of the abusive user in the graph and the overall aspect of the conversation.

Each remaining cluster (49, 119, and 172) focuses on a measure of the \textit{After} graph, highlighting their contribution to class discrimination. We examine more thoroughly these features, by considering separately their distributions in the \textit{Abuse} and \textit{Non-abuse} classes. For the number of users in the conversation, it turns out these distributions are quite different: the Vertex Count is relatively homogeneous and centered around $40$ for the \textit{Abuse} class, whereas its distribution is heterogeneous (closer to a power law) when there is no abuse, with a very large number of very small networks (less than $5$ users).

Looking at the Reciprocity, there is again a relatively homogeneous distribution for the \textit{Abuse} class, centered around $0.7$. For the \textit{Non-abuse} class, a part of the distribution is quite similarly homogeneous (albeit around $0.6$), but the large majority of instances have either a $0$ or $1$ Reciprocity, \textit{i.e.} only unilateral or bilateral edges, respectively. After verification, the former case corresponds to conversations that come to an abrupt end in a short-lived conversation channel. The latter case is just a normally functioning conversation, in which every user talks to each other. Both cases are more likely to happen when few users are involved, which is consistent with our observations regarding the Vertex Count. However, both features are only partially correlated, which shows that the \textit{Abuse} class cannot be reduced only to a question of number of users involved in the conversation.

The Closeness seems to have a special role, since its weighted variants constitute their own clusters for the \textit{Before} (Cluster 110) and \textit{After} (Cluster 177) graphs. By comparison, the average \textit{unweighted} Closeness is correlated with many other features as it belongs to the large Cluster 10: this is consistent with our previous observation that certain weighted variants appear to be more informative. Further examination shows that the Closeness follows a power law-like distribution in both classes, covering three orders of magnitude. However, this heterogeneity is much more marked in the case of the \textit{Non-abuse} class. Concretely, the Closeness is generally higher for the \textit{Abuse} class. This means that the average distance between the author of the targeted message and the rest of the graph decreases in case of abuse. This user becomes less peripheral (or more central), and the same goes for the other users of the graph (in average). This fits in quite well with assumptions about how abuse impacts a discussion: an abuser would tend not to be peripheral in a conversation, while we can reasonably assume that the other participants will be piling on and therefore be less peripheral themselves. 

Most mesoscopic measures are discarded during the feature elimination process. The only remaining ones are the Clique Count and the Coreness, which are also the only ones not related to community structure. Yet, we had considered them as promising in Section~\ref{sec:exp-features-dep}, because they are uncorrelated with the others: it turns out the unique information they convey does not help solving this specific classification task. When inspecting the distribution of the modularity measure, we observe that it is overwhelmingly close to zero in both classes. This means that our networks generally do not have any community structure, which explains why the related features are not discriminative (at least for this dataset).

We also have identified and studied the clusters corresponding to the top features of the \textit{Before} (Table~\ref{tab:SubsetBefore}), \textit{After} (Table~\ref{tab:SubsetAfter}), and \textit{Full} (Table~\ref{tab:SubsetFull}) feature sets. For matters of space, we do not present them in detail, though, and only discuss our most interesting observations. 
Certain clusters identified for the \textit{All} feature set also appear for the other sets: those focusing on the considered type of graph, \textit{i.e.} Clusters 110 and 118 for \textit{Before}; 10, 41, 49 and 177 for \textit{After}, and 9, 10, 41 and 49 for \textit{Full}. Some of the missing clusters are replaced by semantically close and relatively correlated clusters. For instance, \textit{Before} has a cluster 
containing exactly the same measure variants as Cluster 49 (Vertex Count and Betweenness), but for the \textit{Before} graph. Similarly, \textit{Full} has a cluster 
focusing on the weighted Closeness, as Cluster 177 does for \textit{After}. We interpret Clusters 9 and 41 as describing the microscopic aspects of the \textit{After} graph at the graph and vertex scale, respectively: \textit{Before} has comparable clusters 
focusing similarly on the \textit{Before} graph. Overall, we can say that when focusing on a specific type of graph (by opposition to \textit{All}), the classifier takes advantage of informationally close clusters, albeit inferior in terms of discriminative power, as they lead to a lower performance.

 \begin{table}[!ht]
 	\centering
     \caption{Clusters containing the top features (in bold) obtained for the \textit{Before} feature set. Same remarks as for Table~\ref{tab:SubsetAll}.}
     \begin{tabular}{|r|l|l|l|l|l|}
         \hline
      	\textbf{Clust.} & \textbf{Measure} & \textbf{Graph} & \textbf{Wght.} & \textbf{Dir.} & \textbf{Scale} \\
         \hline
         110 & \multicolumn{5}{|l|}{Cf. Table~\ref{tab:SubsetAll}} \\
         \hline
 		\multirow{4}{*}{113} & Clique Count & B & -- & -- & G \\
 		 & \textbf{Coreness Centrality} & \textbf{B} & -\textbf{}- & U/I/\textbf{O} & \textbf{G} \\
 		 & Degree Centrality & B & U & U/I/O & G \\
 		 & Strength Centrality & B & W & U/I/O & G \\
 		\hline
 		\multirow{2}{*}{114} & Degree Centrality & B & U & U/I/O & N \\
 		 & \textbf{Strength Centralit}y & \textbf{B} & \textbf{W} & U/I/\textbf{O} & N \\
         \hline
         118 & \multicolumn{5}{|l|}{Cf. Table~\ref{tab:SubsetAll}} \\
 		\hline
 		\multirow{2}{*}{123} & \textbf{Vertex Count} & \textbf{B} & \textbf{--} & \textbf{--} & \textbf{G} \\
 		 & Betweenness Centrality & B & U/W & U/D & G \\
 		\hline
 		131 & \textbf{Eigenvector Centrality} & \textbf{B} & \textbf{U}/W & \textbf{U} & \textbf{N} \\
         \hline
   	\end{tabular}
 	\label{tab:SubsetBefore}
 \end{table}

 \begin{table}[!ht]
 	\centering
     \caption{Clusters containing the top features (in bold) obtained for the \textit{After} feature set. Same remarks as for Table~\ref{tab:SubsetAll}.}
     \begin{tabular}{|r|l|l|l|l|l|}
         \hline
      	\textbf{Clust.} & \textbf{Measure} & \textbf{Graph} & \textbf{Wght.} & \textbf{Dir.} & \textbf{Scale} \\
         \hline
         10 & \multicolumn{5}{|l|}{Cf. Table~\ref{tab:SubsetAll}} \\
 		\hline
 		40 & \textbf{Coreness} & \textbf{A}/F & -- & U/I/\textbf{O} & \textbf{N} \\
         \hline
         41 & \multicolumn{5}{|l|}{Cf. Table~\ref{tab:SubsetAll}} \\
         \hline
         49 & \multicolumn{5}{|l|}{Cf. Table~\ref{tab:SubsetAll}} \\
 		\hline
 		71 & \textbf{Local Transitivity} & \textbf{A}/F & \textbf{U}/W & \textbf{U} & \textbf{N} \\
 		\hline
 		175 & \textbf{Adhesion}/Cohesion & \textbf{A} & \textbf{--} & \textbf{D} & \textbf{G} \\
 		\hline
 		\multirow{2}{*}{176} & \textbf{Closeness} & \textbf{A} & \textbf{U} & \textbf{U}/I & \textbf{N} \\
 		 & Eccentricity & A & U & U/I/O & N \\
          \hline
         177 & \multicolumn{5}{|l|}{Cf. Table~\ref{tab:SubsetAll}} \\
 		\hline
 		191 & \textbf{Closeness} & \textbf{A} & \textbf{W} & \textbf{I} & \textbf{G}/N \\
         \hline
   	\end{tabular}
 	\label{tab:SubsetAfter}
 \end{table}

 \begin{table}[!ht]
 	\centering
     \caption{Clusters containing the top features (in bold) obtained for the \textit{Full} feature set. Same remarks as for Table~\ref{tab:SubsetAll}.}
     \begin{tabular}{|r|l|l|l|l|l|}
         \hline
      	\textbf{Clust.} & \textbf{Measure} & \textbf{Graph} & \textbf{Wght.} & \textbf{Dir.} & \textbf{Scale} \\
         \hline
         9 & \multicolumn{5}{|l|}{Cf. Table~\ref{tab:SubsetAll}} \\
         \hline
         10 & \multicolumn{5}{|l|}{Cf. Table~\ref{tab:SubsetAll}} \\
 		\hline
  		39 & \textbf{Closeness} & \textbf{F} & \textbf{W} & \textbf{U} & G/\textbf{N} \\
         \hline
         41 & \multicolumn{5}{|l|}{Cf. Table~\ref{tab:SubsetAll}} \\
         \hline
         49 & \multicolumn{5}{|l|}{Cf. Table~\ref{tab:SubsetAll}} \\
 		\hline
   	\end{tabular}
 	\label{tab:SubsetFull}
 \end{table}

\subsection{Baseline Comparison}
\label{sec:res-baseline}
For matters of exhaustiveness, we assess the performance of our framework on a \textit{balanced} version of our classes, instead of the unbalanced ones used throughout this section. In this setting, the \textit{Abuse} class stays the same, but the \textit{Non-abuse} class is reduced to the size of the \textit{Abuse} one, through sampling. When using these data, we observe a significant performance improvement for all feature sets. In particular, the $F$-measure values obtained for \textit{All} and \textit{All-TF} increase from $83.99$ and $82.65$, to $88.87$ and $87.10$, respectively. Further investigations shows that this improvement is mainly due to a decrease in the number of false positives, itself caused by the smaller size of the non-abuse class. Such a balanced situation is unlikely in practice, though.

\begin{table}[!ht]
\centering
\caption{Best performances for the baselines and current framework.}
  \begin{tabular} { |l|l|r| }
    \hline
    \textbf{Method} & \textbf{Reference} & \textbf{$F$-measure} \\ 
    \hline
    Content-based & \cite{Papegnies2017a} & 76.50 \\
    Graph-based & \cite{Papegnies2017} & 77.00 \\
    Extended graph-based -- \textit{All} & This article & \textbf{83.89} \\
    Extended graph-based -- \textit{All-TF} & This article & 82.65 \\
    \hline
  \end{tabular}
  \label{tab:BaselinePerfs}
\end{table}

Finally, we compare the results obtained using our framework with our two baselines (Table~\ref{tab:BaselinePerfs}): the content-based approach of \cite{Papegnies2017a} and the previous version of our graph-based method \cite{Papegnies2017}. As a reminder, the main differences between the latter and our present framework are that we now extract a directed graph, and use a much larger number of topological measures as classification features. The combination of these two improvements leads to a significant performance increase over our previous effort. As described in Section~\ref{sec:res-weigths-direction}, the contribution of edge directions to the overall performance is relatively minor. One could assume that the performance improvement is mainly caused by the major expansion of the feature set, however this improvement is observed even when only using the top features identified in the previous section. Yet, there are only $10$ of them, by comparison to the $75$ features used in our previous approach~\cite{Papegnies2017}. So the conclusion here is that both extracting a directed graph and selecting a more appropriate set of features (in particular, topological measures able to handle edge weights) helped improve the performance. More importantly, the performance is greatly improved compared to our content-based approach \cite{Papegnies2017a}, which is quite representative of the preprocessing and features used in the literature when classifying such data. This is a major result, as it shows that the sole structure of the conversation is enough to efficiently detect abuses, without considering at all the content of the exchanged messages.

\section{Conclusion}
\label{sec:Conclusion}
In this article, we tackle the problem of automatic abuse detection in online communities. We propose to model online conversations through graphs, and to perform the classification task using only graph-based features. The method, while simple, yields good results (up to a $83.89$ $F$-measure), besting the score obtained with a content-based approach~\cite{Papegnies2017a} and our previous graph-based effort~\cite{Papegnies2017}. It completely ignores the content of messages exchanged between users of an online community, which means it is robust to intentional obfuscation of messages by abusive users, as well as unintentional content noise. It is also inherently language-independent. One important limitation of our method is the high computational time required to extract the features. However, we show that it can be very significantly reduced by working with a small subset of relevant features, resulting in more than $97\%$ of the original performance for less than $0.01\permil$ of the processing time. We also show that while our method is originally not designed for real-time abuse detection, the information available at the time the message appears is discriminative enough to do so.

A straightforward extension of our work is to take advantage of both content- and graph-based features, an approach previously applied in other contexts~\cite{Rumshisky2017}. In our case, they are both based on completely different types of information, so we can assume they are complementary, which could improve the classification performance. At the very least, it will be interesting to combine the features of the \textit{Before} graph with textual features since that can lead to a system useful for a prediction task. We also consider using a content-based classifier in a completely different way, during the graph extraction process. Such a classifier could be trained to detect the nature of the interaction between two users, allowing to extract a signed network (negative edge for a hostile exchange, positive otherwise). This additional information is likely to improve the performance of our graph-based classifier.

Finally, part of the future work will focus on applying our proposed approaches to other types of social network corpora. Indeed, chat logs are a special case of communications records that have a very specific structure (entanglement of discussions, near-synchronous communications, various topics in a single flow of discussion, uncertainty about who is the intended recipient of a message...) which do not necessarily appear in other forms of social networks, such as forums or micro-blogs.
For example, since our results have shown that directionality is not the dominant graph construction parameter, we would be interested in evaluating its impact on a type of social media integrating a clear response structure (\textit{i.e.} a clear identification of who answers whom, such as in a forum like Reddit or a tree-shaped comment section of a news website). Additionally, another type of social network corpora might present a more distinct community structure and therefore render the meso-scale features we have presented more relevant.



\section*{Acknowledgments.}
This work was funded by a grant from the \textit{Provence-Alpes-C\^ote-d'Azur} region (France) and the \textit{Nectar de Code} company.

\bibliographystyle{IEEEtran}
\bibliography{comp}

\newpage
\begin{IEEEbiography}[{\includegraphics[width=1in,height=1.25in,clip,keepaspectratio]{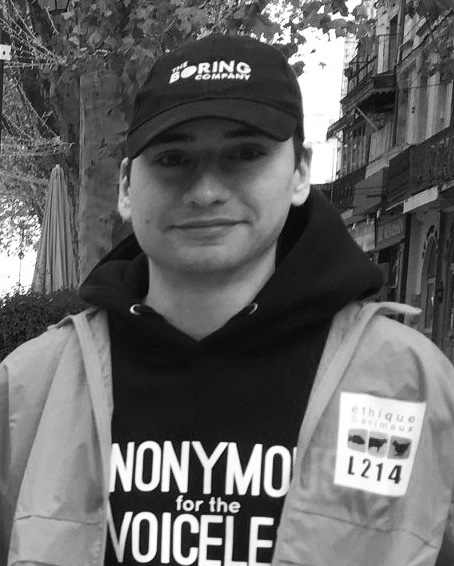}}]{Etienne Papegnies}
Etienne Papegnies received License (2013) and Master (2015) degrees in Software Engineering from Avignon University (France). He has worked on the analysis of EEG (Electroencephalography) signal in the context of human machine interface. His Master thesis focused on approximate text searching in noisy documents. In 2015, he started a PhD in Computer Science also at Avignon Universit\'e, in the LIA (Computer Science laboratory), on the development of automatic moderation methods for social media. This PhD is funded by both the \textit{Nectar of Code} company and the \textit{Provence-Alpes-Côtes d'Azur} region (France).
\end{IEEEbiography}

\vspace{-2.5cm}
\begin{IEEEbiography}[{\includegraphics[width=1in,height=1.25in,clip,keepaspectratio]{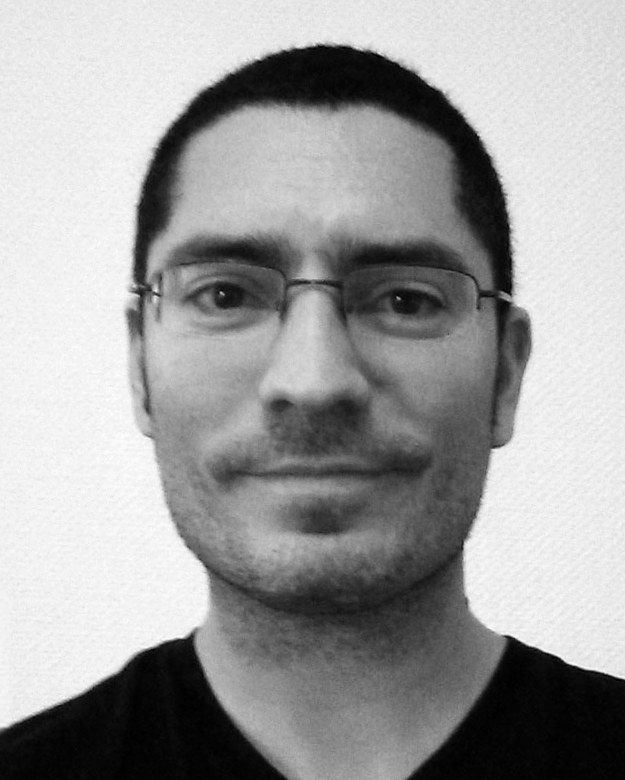}}]{Vincent Labatut}
Vincent Labatut received his PhD in Computer Science / Artificial Intelligence from the Universit\'e Paul-Sabatier Toulouse III (France) in 2003. His work focused on defining a paradigm allowing to model cerebral information and its processing, and took place at INSERM (France). Between 2003 and 2005 he was a lecturer at the Universit\'e Paul-Sabatier Toulouse III and co-founded Personnalit\'e Num\'erique SAS, a data storage start-up. From 2005 to 2014, he was an assistant professor at the Galatasaray University (Istanbul, Turkey). Since 2014, he has been an associate professor at the Computer Science Laboratory (LIA) of Avignon Universit\'e (France). His current research interests include complex network analysis (centrality, community detection, signed networks, etc.), as well as information retrieval (especially on graphs).
\end{IEEEbiography}

\vspace{-2.5cm}
\begin{IEEEbiography}[{\includegraphics[width=1in,height=1.25in,clip,keepaspectratio]{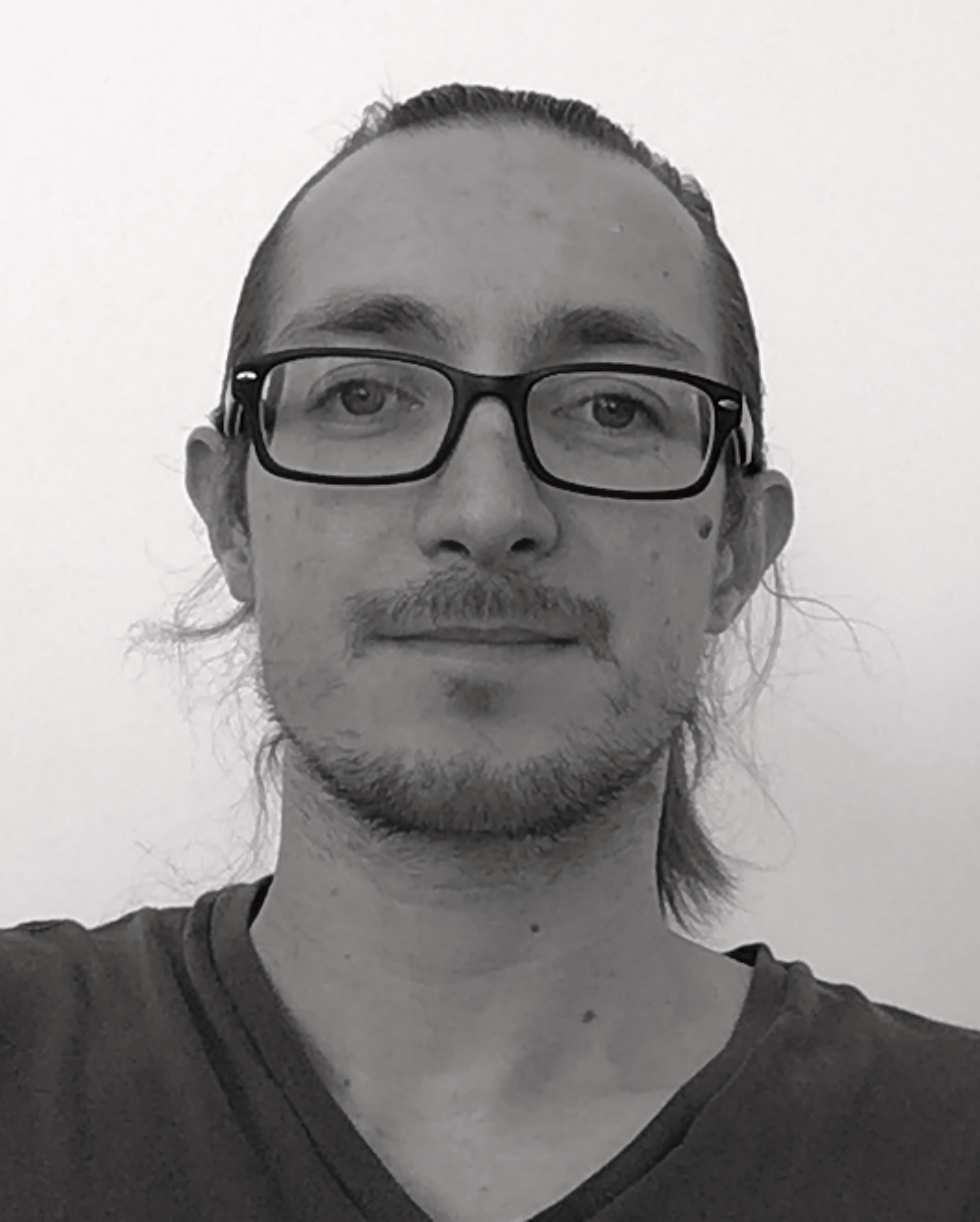}}]{Richard Dufour}
Richard Dufour received the M.S. and Ph.D. degrees from Le Mans University (France), in 2007 and 2010 respectively. Since 2012, he is an associate professor in computer science at LIA laboratory within Avignon University (France) where he is part of the speech and language processing team. His research interests include speech and language processing, and extraction as well as analysis of textual documents in noisy conditions. He is a member of several funded projects where he is working on document classification, on speech processing, and on information representation from social media.
\end{IEEEbiography}

\vspace{-2.5cm}
\begin{IEEEbiography}[{\includegraphics[width=1in,height=1.25in,clip,keepaspectratio]{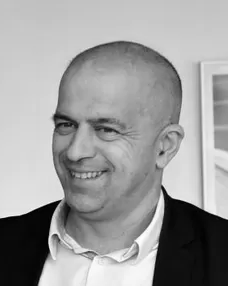}}]{Georges Linar\`{e}s}
Georges Linar\`{e}s has been a full professor in computer science since 2011 at Avignon Universit\'e. He graduated in mathematics in 1995 before obtaining a PhD in the field of neural networks for automatic classification in 1998 and joining the Speech Processing group of the Computer
Science Laboratory of Avignon (LIA). His main research interests are related to natural language processing, multimedia information retrieval and machine learning. He supervised 15 PhD students and authored or co-authored about 150 articles in the main journals and conferences of these research fields. He coordinated the \textit{Neurocomputation and Language Processing} group of the \textit{Brain and language Research Institute} (France) and headed the LIA from 2010 to 2015. He is currently pro-vice-chancellor for research at Avignon Universit\'e.
\end{IEEEbiography}
\enlargethispage{-3cm}

\end{document}